\documentclass[aps,pre,reprint,floatfix,groupedaddress]{revtex4-1}

\usepackage{graphicx}
\usepackage[utf8]{inputenc}

\usepackage{amsmath}
\usepackage{amssymb}
\usepackage{bm}
\usepackage{cancel}
\usepackage[normalem]{ulem}
\usepackage{xfrac}

\usepackage{textcomp}

\usepackage{color}

\usepackage{hyperref}
\usepackage{cleveref}

\usepackage{subfigure}
\usepackage{placeins}

\newcommand{\sgn}{\text{sgn}}
\newcommand{\eq}{\text{eq}}

\newcommand{\calH}{\mathcal{H}}

\newcommand{\nose}{\text{n}}

\newcommand{\fla}{\text{L}}

\newcommand{\dfb}{\delta\! f_{2,b}}
\newcommand{\dfc}{\delta\! f_{2,c}}

\newcommand{\be}{\begin{equation}}
\newcommand{\ee}{\end{equation}}

\begin{document}

\title{Bifurcation analysis and phase diagram of a spin-string model
  with buckled states}
%

\author{M.~Ruiz-Garcia}
\email{miruizg@ing.uc3m.es}

\author{L.~L.~Bonilla} 

\affiliation{$^1$Gregorio Mill\'an Institute for Fluid Dynamics, Nanoscience, and Industrial Mathematics, and Department of Materials Science and Engineering and Chemical Engineering, Universidad Carlos III de Madrid, Avenida de la Universidad 30, 28911 Legan\'es, Spain}

\author{A.~Prados$^{2}$}

\affiliation{$^2$F\'{\i}sica Te\'{o}rica, Universidad de Sevilla,
Apartado de Correos 1065, E-41080, Sevilla, Spain}

\date{\today}

\begin{abstract}
  We analyze a one-dimensional spin-string model, in which string
  oscillators are linearly coupled to their two nearest
  neighbors and to Ising spins representing internal degrees of freedom. String-spin coupling induces a long-range ferromagnetic interaction among spins that competes with a spin-spin antiferromagnetic coupling.  As a
  consequence, the complex phase diagram of the system exhibits different flat rippled and
  buckled states, with first or second order transition lines between states. The two-dimensional version of the model has a similar phase diagram,
  which has been recently used to explain the rippled to buckled
  transition observed in scanning tunnelling microscopy experiments
  with suspended graphene sheets. Here we describe in detail the phase diagram of the simpler one-dimensional model and phase stability using bifurcation theory. This
  gives additional insight into the physical mechanisms underlying the
  different phases and the behavior observed in experiments.
\end{abstract}

\maketitle

\section{Introduction}

Rippling and buckling of suspended graphene sheets is an active
research topic
\cite{mey07,fas07,gazit09,san-jose,gonzalez,prb12bon,jsm12bon,schoelz15,ByR16}. Recent
scanning tunnelling microscopy (STM) experiments show that local
heating induces a transition from a soft rippled sheet to a hard
buckled graphene membrane \cite{schoelz15}. While heating certainly
increases thermal fluctuations, in this case and quite
counterintuitively, it produces a more ordered phase. We have
interpreted the STM experiments as the result of driving the system
through a first order phase transition between flat and buckled
membrane states \cite{RByP16}. We have used a phenomenological
spin-membrane model of the graphene sheet that exhibits such a first
order phase transition \cite{RByP16}, for there is no first principles
derivation thereof. More specifically, spin-string \cite{pre12bon} and
spin-membrane models \cite{jsm12bon,Ruiz-Garcia,RByP16} display first
and second order buckling transitions depending on the precise
interactions among spins. In fact, spin-spin interactions and
spin-membrane are reminiscent of the effective interactions among out
of plane displacements in 2d systems, which appear after electrons and
in plane phonons are integrated out in electron-phonon models
\cite{san-jose,gui14}. Also, mechanical systems coupled to
  spins have been employed to describe structural phase transitions in
  other physical contexts
  \cite{Ka60,Py73,FyP73,Ri77,Ri80,SyS73,SyS77,PTyZ07,PTyZ08,jstat10,jstat10a,PCyB12}.

In these models, the membrane is a system of mass points on a lattice
that move vertically and are interconnected by linear springs. There
is a pseudo-spin at each lattice node, which represents in a simple
way some internal degrees of freedom, that pushes the point mass
located there either upwards or downwards. If the pseudo-spins are
coupled only to mass points but not among themselves, there is a
second order buckling transition below a critical temperature
\cite{pre12bon}. This is also the case for a membrane described by
F\"oppl-von K\'arm\'an equations on a hexagonal lattice with vertical
displacements coupled to the local spin on the same lattice node
\cite{jsm12bon}. This second order transition arises because the
spin-membrane brings about a long-ranged ferromagnetic interaction
among the pseudo-spins. Furthermore, additional short range
antiferromagnetic couplings among the pseudo-spins produce different
phases and first or second order transitions among
them \cite{Ruiz-Garcia,RByP16}.

One drawback of the two-dimensional ($2d$) spin-membrane models with
antiferromagnetic coupling is that most results are obtained from
numerical simulations. In this work, we study analytically the
corresponding one-dimensional ($1d$) spin-string model, in terms of
the dimensionless temperature $\theta$ and the spin-spin
antiferromagnetic coupling $\kappa$. As already said above, for
$\kappa=0$ there is a second order phase transition at $\theta=1$ from a flat string configuration (stable for $\theta>1$) to stable buckled string states that exist for $\theta<1$. This second order phase transition is a supercritical pitchfork bifurcation \cite{Io90}. For $\kappa\neq 0$, we find and analyze subcritical pitchfork bifurcations corresponding to first order phase transitions between flat and buckled phases. This situation is similar to the $2d$ case but, in $1d$, we are able to obtain bifurcation lines, bifurcation diagrams and the different phases by analytical methods. The order parameter spin magnetization acts as the norm of the solution in bifurcation diagrams \cite{Io90}.

  We show that the flat string configuration is the only stable phase except for a finite region within the first quadrant $\kappa>0$, $\theta>0$ of the $(\kappa,\theta)$ plane. The flat string configuration is unstable inside a smaller region $\kappa\in (0,\kappa_\nose)$, $\theta\in(0,1)$, which is bounded by a two-valued curve $\theta_b(\kappa)$ joining the origin to $(\kappa,\theta)=(0,1)$. This curve is a locus of pitchfork bifurcations from flat to buckled states. Bifurcations are subcritical at the low $\theta$ branch of the curve. At the high branch, they are supercritical for $0<\kappa<\kappa_c$ and subcritical for $\kappa_c<\kappa<\kappa_\nose$. The tricritical point \cite{La80,Ye93} $(\kappa_c,\theta_c)$ has codimension two and we can use two-parameter perturbation theory to analyze the change from super to subcritical pitchfork bifurcation. The result is that the branch of unstable buckled states that stem from the flat configuration for $\theta<\theta_c$ coalesce with a branch of stable buckled states at a curve $\theta_M(\kappa)$ that is above the bifurcation curve $\theta_b(\kappa)$. This signals a first order phase transition and bistability between flat and buckled states. As the low and high temperature branches of $\theta_b(\kappa)$ coalesce at the turning point $(\kappa_\nose,\theta_\nose)$, the corresponding subcritical bifurcations merge and disappear. Analysis of this new codimension two point shows that there exist an isola of buckled states with positive magnetization that is not connected to the flat string configuration (a symmetric isola with negative magnetization also exists). For fixed values of $\kappa$ and $\theta$, there are two buckled states: that with larger (smaller) magnetization is stable (unstable). These two buckled states coalesce for a sufficiently large temperature at the curve $\theta_M(\kappa)$.

The first order phase transition occurring in this $1d$ spin-string
model is akin to that found numerically in the $2d$ spin-membrane
model. Our explanation of Schoelz {\em et al}'s experiments
\cite{schoelz15} is that the STM drives the system dynamically across
the first order phase transition appearing in a certain range of
antiferromagnetic coupling \cite{RByP16}. The same situation occurs in the $1d$ spin-string model.

The paper is organized as follows. In Sec.~\ref{sec:1dmodel}, we
define the model and introduce the free energy density controlling its
equilibrium behavior, together with the corresponding Euler-Lagrange
equation governing the equilibrium profiles. Also, we briefly discuss
the flat solution and its stability. Section \ref{sec:phase-diagram}
puts forward the main results of our study, including a discussion of
main elements of the phase diagram of the system, leaving the
derivations for the later sections. We analyze in detail the
bifurcation from the flat solution in Sec.~\ref{sec:bifurcation} and
the emergence of a (tri)critical point, at which the transition
changes from second-order to first-order. In
Sec.~\ref{sec:exact-sols}, we study the low temperature limit of our
system, focusing on the spin configurations underlying the parabolic
profiles of the string.  We present the main conclusions of our work
in Sec.~\ref{sec:conclusions}.  The appendices deal with some
technical details and calculations that are omitted in the main text.

\section{Continuum limit of the spin-string model} \label{sec:1dmodel}

We consider a spin-string system with Hamiltonian
\begin{equation}
 \calH(\bm{u},\bm{p},\bm{\sigma})\!=\!\sum_{j=0}^{N} \biggl[ \frac{p_j^2}{2m}+ \frac{k}{2}(u_{j+1}-u_j)^2-fu_j\sigma_j
 + J \sigma_{j+1}\sigma_j\biggr].
\label{H1}
\end{equation}
Here, $u_{j}$ and $p_{j}$, $j=1,\ldots,N$, are the string vertical
displacements and their conjugate momenta, respectively, and
$\sigma_{j}=\pm 1$ are pseudo-spin variables
\cite{pre12bon,Ruiz-Garcia}. The latter represent internal degrees of
freedom arising from internal forces that push the atoms along the
vertical direction. Therefore, we have: (i) a nearest-neighbor
harmonic interaction between the elastic variables,
$k(u_{j+1}-u_{j})^{2}$, (ii) an on-site interaction between the
elastic and the internal variables, $-fu_{j}\sigma_{j}$, and (iii) a
nearest-neighbor spin-spin interaction, $J \sigma_{j+1}\sigma_j$. We
have clamped boundary conditions at the string ends,
$u_0=p_0=\sigma_0=u_{N+1}=p_{N+1}=\sigma_{N+1}=0$.

The string variables $u_{j}$ and $p_j$ satisfy Hamilton's equations of
motion, whereas the pseudo-spins $\sigma_{j}$ evolve following Glauber
dynamics \cite{Gl63} at the thermal bath temperature $T$
\cite{Ruiz-Garcia}. Then the system reaches equilibrium in the long
time limit. The probability density of finding the system in a certain
configuration $(\bm{u},\bm{p},\bm{\sigma})$ is given by
$e^{-\mathcal{H}/T}/Z$, where $Z$ is the partition function and we
have set $k_{B}=1$.  For $J=0$ and temperature below
$T_0=\frac{f^2 N^{2}}{k\pi^{2}}$, the system exhibits stable ripples \cite{pre12bon}. We now make energy variables dimensionless by
measuring them in units of $T_{0}$. The dimensionless coupling
constant $\kappa$ and temperature $\theta$ are
\begin{equation}
  \kappa=\frac{J}{ T_{0}}, \quad
  \theta=\frac{T}{T_{0}}.
\label{par2}
\end{equation}
Suitable units are introduced for the remaining variables. Further
details can be found in Ref.~\cite{Ruiz-Garcia}.

In this paper, we investigate the equilibrium states and the different
phases of the model in the limit as $N\gg 1$ with $x=i/N\in [0,1]$
\cite{Ruiz-Garcia}.  We integrate out the pseudo-spins and the
canonical momenta. Then the resulting equilibrium probability density
$\mathcal{P}[u]$ of finding the string with a certain profile $u(x)$
is
$\mathcal{P}[u;\theta,\kappa]
\propto\exp{\left(-F[u;\theta,\kappa]/{\theta}\right)}$, in which
\begin{subequations}\label{eq:A-and-ln-zeta}
\begin{equation}\label{eq:A-of-u}
 F[u;\theta,\kappa]=N \int_{0}^{1} dx \,
 f(u,u';\theta,\kappa),
\end{equation}
\begin{equation}
 f(u,u';\theta,\kappa)= \frac{(u^{\prime})^{2}}{2\pi^{2}}-
\theta \ln\zeta\left(
\frac{u}{\theta},\frac{\kappa}{\theta} \right),
\end{equation}
\begin{eqnarray}
\zeta\!\left(
\frac{u}{\theta},\frac{\kappa}{\theta} \right)\!&=&\exp\!\left(-\frac{\kappa}{\theta}\right)
         \cosh\!\left(\frac{u}{\theta}\right) \nonumber \\
&+&\!\exp\!\left(\frac{\kappa}{\theta}
\right)\!
   \sqrt{1+\exp\!\left(-\frac{4\kappa}{\theta}\right)\!\sinh^{2}\!\left(\frac{u}{\theta}\right)}. 
   \label{eq:ln-zeta}
\end{eqnarray}
\end{subequations}
In the equations above, $F[u;\theta,\kappa]$, $f(u,u';\theta,\kappa)$,
and $\ln\zeta(u;\theta,\kappa)$ are the total free energy, the (local)
free energy density per unit length, and the logarithm of the
pseudo-spins partition function per site, respectively.

\subsection{Euler-Lagrange equation for the equilibrium profiles}\label{sec:Euler-Lagrange}

The equilibrium profiles $u_{\eq}(x)$ solve the Euler-Lagrange equation,
\begin{equation}\label{eq:Euler-Lagrange}
 \frac{1}{\pi^{2}}u_{\eq}''=-\mu(u_{\eq};\theta,\kappa), \qquad u_{\eq}(0)=u_{\eq}(1)=0.
\end{equation}
where
\begin{equation}\label{eq:local-magnetization} \mu(u;\theta;\kappa)\!\equiv\!-\frac{\partial f(u,u';\theta,\kappa)}{\partial u}\!=\!\frac{e^{-\frac{2 \kappa}{\theta}} \sinh \left(\frac{u}{ \theta}\right)}{\sqrt{e^{-\frac{4 \kappa}{\theta}}\! \sinh^{2} \left(\frac{u}{ \theta}\right)+1}},
\end{equation}
is the local value of the magnetization. Clearly, the magnetization sets the local value
of the string curvature.

Let us consider only the first buckled mode that has no internal nodes. The absolute value of the total magnetization distinguishes between buckled and flat profiles and it is therefore an order parameter
\begin{equation}\label{eq:magn}
 M(\theta;\kappa)=\left|\,\int_{0}^{1} dx \, \mu(u_{\eq};\theta,\kappa) \,\right|.
\end{equation}
Further information is given by the parameter
\begin{equation}\label{eq:DL} \mathcal{DL}(\theta;\kappa)=\frac{1}{2}\left(1-\theta\int_{0}^{1}dx\,\left.\frac{\partial\ln\zeta}{\partial\kappa}\right|_{u=u_{\eq}}\right),
\end{equation}
which is zero for perfect anti-ferromagnetic order, 1/2 for a random configuration of the pseudo-spins, and 1 for perfect ferromagnetic order \cite{Ruiz-Garcia}. Recall that $C=-\theta\, \partial_{\kappa} \ln\zeta$ gives the correlation of nearest-neighbor pseudo-spins.

The free energy functional $F[u]$ has a relative (or weak) minimum for the curve $u=u_{\eq}(x)$ provided the following two conditions are satisfied:
\begin{enumerate}
\item the curve $u_{\eq}(x)$ must satisfy the Euler-Lagrange equation
  \eqref{eq:Euler-Lagrange}.
\item The linearised Euler-Lagrange equation about $u_{\eq}(x)$,
\begin{equation}\label{eq:EL-linearized}
\qquad \delta u''=-\pi^{2}\left(\frac{\partial \mu}{\partial
    u}\right)_{u=u_{\eq}}\!\delta u, \,\, \delta u(0)=\delta u(a)=0,
\end{equation}
must have only the trivial solution $\delta u(x)\equiv 0$, $\forall
x$,  for any $a\leq 1$.
\end{enumerate}
Considered separately, each condition is necessary for $F[u]$ to have
a weak minimum (with the nuance $a<1$ instead of $a\leq 1$ in the
second one) \cite{Gelfand-Fomin}.

\subsection{Flat string profile and its stability}
\label{sec:flat-profile}

The flat string profile $u_{\fla}(x)\equiv 0$, $\forall x$, is always a solution of the Euler-Lagrange equation, which we call phase L \cite{note2}. It is (locally) stable if it corresponds to a minimum of the free energy functional. For phase L and any $a\leq 1$, the boundary value problem~\eqref{eq:EL-linearized} is
\begin{equation}\label{eq:EL-flat-lin}
\delta u''=-\pi^{2}\,\theta^{-1}\exp(-2\kappa/\theta)\,\delta u,
\, u(0)=u(a)=0.
\end{equation}
Aside from the trivial solution $\delta u(x)\equiv 0$, we may have solutions
\begin{equation}\label{eq:sin-profiles}
\delta u(x)= A \sin \left[\pi\theta^{-1/2}\exp(-\kappa/\theta)x\right],
\end{equation}
where $A$ is an arbitrary constant and $a$ is such that
\begin{equation}\label{eq:conjugate-point}
\theta^{-1/2}\exp(-\kappa/\theta)a=n, \qquad n\in\mathbb{N}, \; a\leq 1.
\end{equation}

Thus, first, the the flat solution produces a relative minimum of the
free energy if $\theta^{-1/2}\exp(-\kappa/\theta)<1$. In this region
of the $(\kappa,\theta)$ plane, the only solution of
\eqref{eq:EL-flat-lin} is the trivial one.  Second, if
$\theta^{-1/2}\exp(-\kappa/\theta)>1$, there is at least one
nontrivial solution of \eqref{eq:EL-flat-lin}, provided we choose
$a=\theta^{1/2}\exp(\kappa/\theta)<1$ and the flat profile is no
longer stable.

Buckled equilibrium profiles may bifurcate at the curve
$\theta^{1/2}\exp(\kappa/\theta)=1$, which is a bifurcation line in
the $(\kappa,\theta)$ plane enclosing Region II in
Fig.~\ref{fig:bifurc}. Points $(\kappa_{b},\theta_{b})$ on this line
satisfy
\begin{equation}\label{eq:bifurc-line}
\theta_{b}\exp\left(\frac{2\kappa_{b}}{\theta_{b}}\right)=1, \text{ or }
\kappa_{b}=-\frac{1}{2}\theta_{b}\ln\theta_{b}.
\end{equation}
The bifurcation line has two branches $\theta^{(2)}(\kappa)<\theta^{(1)}(\kappa)$ that coalesce at the turning point (``nose'') $N\equiv(\kappa_{\nose}=(2e)^{-1},\theta_{\nose}=e^{-1})$,
$\theta_{\nose}=2\kappa_{\nose}$.  For $\kappa>\kappa_{\nose}$, the
free energy has a local minimum at the flat solution, regardless of
the temperature. For $\kappa<\kappa_{\nose}$ the flat solution is
unstable if $\theta^{(2)}(\kappa)<\theta<\theta^{(1)}(\kappa)$, and
locally stable otherwise, see Fig.~\ref{fig:bifurc}. Note that
$\theta^{(2)}<2\kappa<\theta^{(1)}$.  The tangent to the bifurcation line at
$(\kappa_{b},\theta_{b})$ verifies
\begin{equation}\label{eq:bif-tangent}
2\,\delta\kappa_{b}+(1+\ln\theta_{b})\delta\theta_{b}=0.
\end{equation}
Here, $\delta\theta_{b}$ and $\delta\kappa_{b}$ are the (small)
deviations from $(\kappa_{b},\theta_{b})$ over the tangent.

\section{Results: Phase diagram}\label{sec:phase-diagram}

\begin{figure}
\centering
  \includegraphics[width=3.25in]{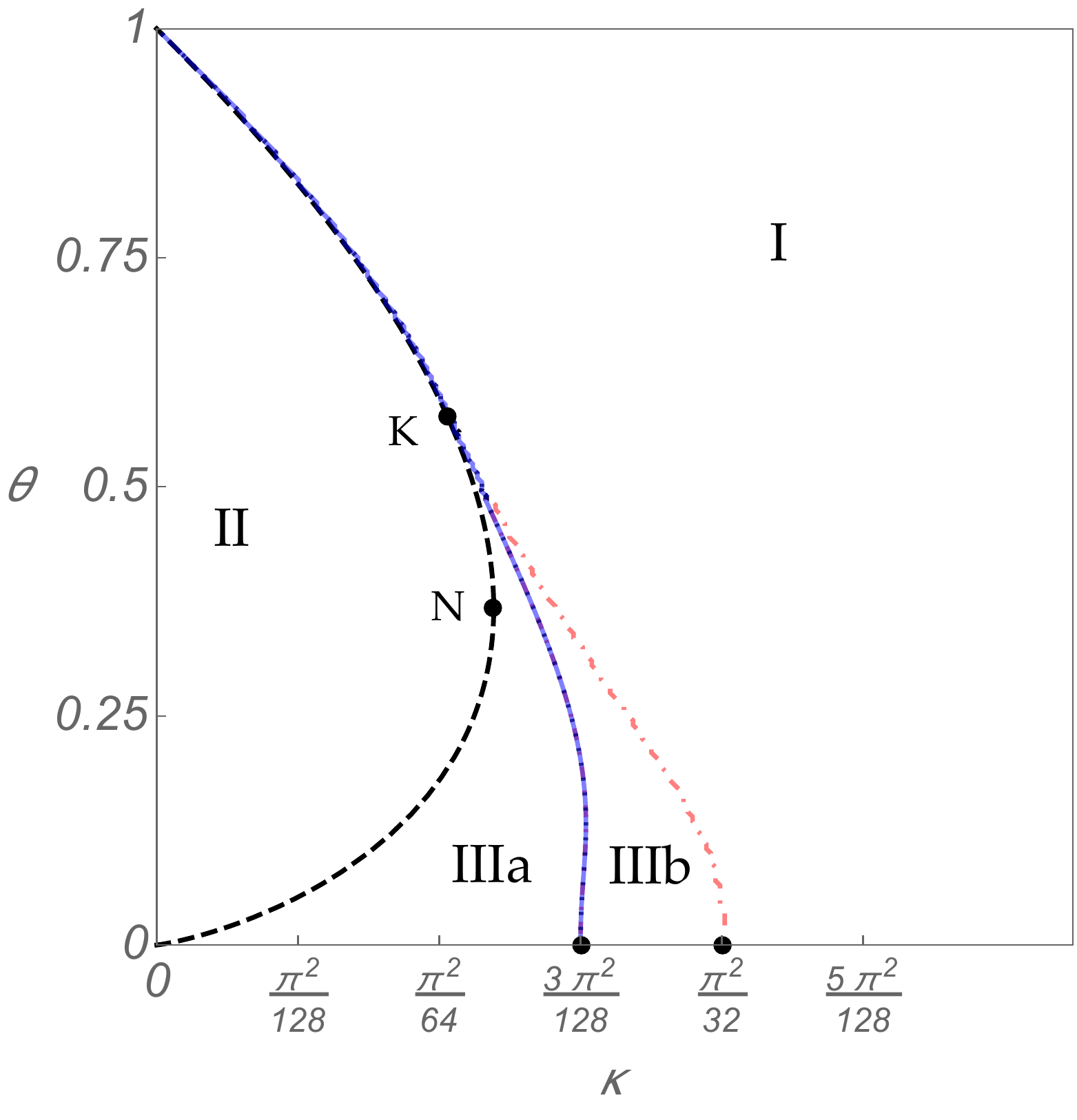}
  \caption{\label{fig:bifurc} Phase diagram in the $(\kappa,\theta)$
    plane. We have marked the tricritical point $K$ and the turning
    point $N$ over the bifurcation curve $\kappa_{b}(\theta)$ (dashed
    line). Also plotted are the coexistence line $\kappa_{t}(\theta)$
    (solid line) and the first-order line $\kappa_{M}(\theta)$ (dotted
    line). Note that
    $\kappa_{b}(\theta)<\kappa_{t}(\theta)<\kappa_{M}(\theta)$. The
    definition of the different regions I, II and III if the phase
    diagram, as well as the existing phases in each region and their
    stability is summarised in Table~\ref{tab:phases}.  In addition,
    the values of $\kappa$ controlling the low temperature behavior,
    $\kappa_{t}^{(0)}=3\pi^{2}/128$ and $\kappa_{M}^{(0)}=\pi^{2}/32$,
    are shown with points.  }
\end{figure}
\begin{table*}
  \centering
  \begin{tabular}{|c|c|c|c|c|c|}
    \hline
    Region   & Definition & Phases & Most stable  & Unstable & Metastable \\
    \hline \hline
    I & \begin{tabular}{cc}
$\kappa>\kappa_{b}(\theta)$ &
    $\theta>\theta_c$ \\
\hline
$\kappa>\kappa_{M}(\theta)$ &
    $\theta<\theta_c$
\end{tabular} & L & L & None & None \\
    \hline
    II & $\kappa<\kappa_b(\theta)$ & B+,L & B+ & L & None \\
    \hline
    IIIa & $\kappa_{b}(\theta)<\kappa<\kappa_{t}(\theta)$ & B+,B-,L
                                   & B+ &  B- & L \\
    \hline
    IIIb & $\kappa_{t}(\theta)<\kappa<\kappa_{M}(\theta)$ & B+,B-,L
                                   & L & B- & B+ \\ \hline                                                                 \end{tabular}
                                 \caption{Summary of the different regions, phases (flat $L$, stable
                                   buckled $B+$ and unstable buckled $B-$) and their relative stability.}
  \label{tab:phases}
\end{table*}


This section describes the main results of this paper, leaving
derivations for later sections. There are three different phases in
the system: the flat phase $L$ and two buckled phases, which we denote
$B+$ and $B-$ (for low temperatures they are the string profiles shown in Fig. \ref{fig:low-temp-profiles}). The points and lines governing the existence and
stability of the different phases are shown in
Fig.~\ref{fig:bifurc}. It shows the bifurcation line $\kappa_b(\theta)$ and the turning point $N$ that separates its two branches $\theta^{(1)}(\kappa)\geq \theta_{\nose}$ and $\theta^{(2)}(\kappa)\leq \theta_{\nose}$. The interior of the bifurcation curve is Region II.

We shall now anticipate some results that will be discussed in depth in Section \ref{sec:bifurcation}.  A key element in the phase diagram is the existence of a tricritical
point $K$, $K\equiv(\kappa_{c}=\sqrt{3}\ln 3/12,\theta_{c}=1/\sqrt{3})$,
at which the three phases $L$, $B+$ and $B-$ coalesce \cite{La80,Ye93}. For $\theta>\theta_{c}$,
the bifurcation at $\theta^{(1)}(\kappa)$ is supercritical, a stable
buckled profile $B+$ stems continuously from the flat solution in region II. For $\theta<\theta_{c}$, the bifurcation becomes subcritical. Then an unstable
buckled profile $B-$ issues from the flat solution at $\theta>\theta^{(1)}(\kappa)$ (upper bifurcation  branch) and and at $\theta<\theta^{(2)}(\kappa)$ (lower branch). The stable buckled
phase $B+$ does not disappear at $K$. Instead, $B+$ and the unstable state $B-$ coalesce at a temperature $\theta_M(\kappa)$ (dotted red line in Fig.~\ref{fig:bifurc}) higher than $\theta^{(1)}(\kappa)$ for $\kappa>\kappa_c$. The transition at $K$ changes to first order. The phase $B+$
exists inside the bifurcation curve (region II) and also outside it (region III). For $\kappa>\kappa_{M}(\theta)$, we have only the flat phase $L$. In region III, there are three phases: $B-$ is unstable,
whereas phases $L$ and $B+$ are both locally stable as they correspond to local minima of the free energy. Their relative
stability depends on $\kappa$: in fact, there appears a coexistence
line $\kappa_{t}(\theta)$ (solid blue in Fig.~\ref{fig:bifurc}) at
which both phases are equiprobable. In region IIIa,
$\kappa_{b}<\kappa<\kappa_{t}$, phase $B+$ provides the absolute minimum
and phase $L$ is metastable, while in region IIIb,
$\kappa_{t}<\kappa<\kappa_{M}$, the situation is reversed.

\begin{figure}
	\centering
	\subfigure[]{\includegraphics[width=1.6125in]{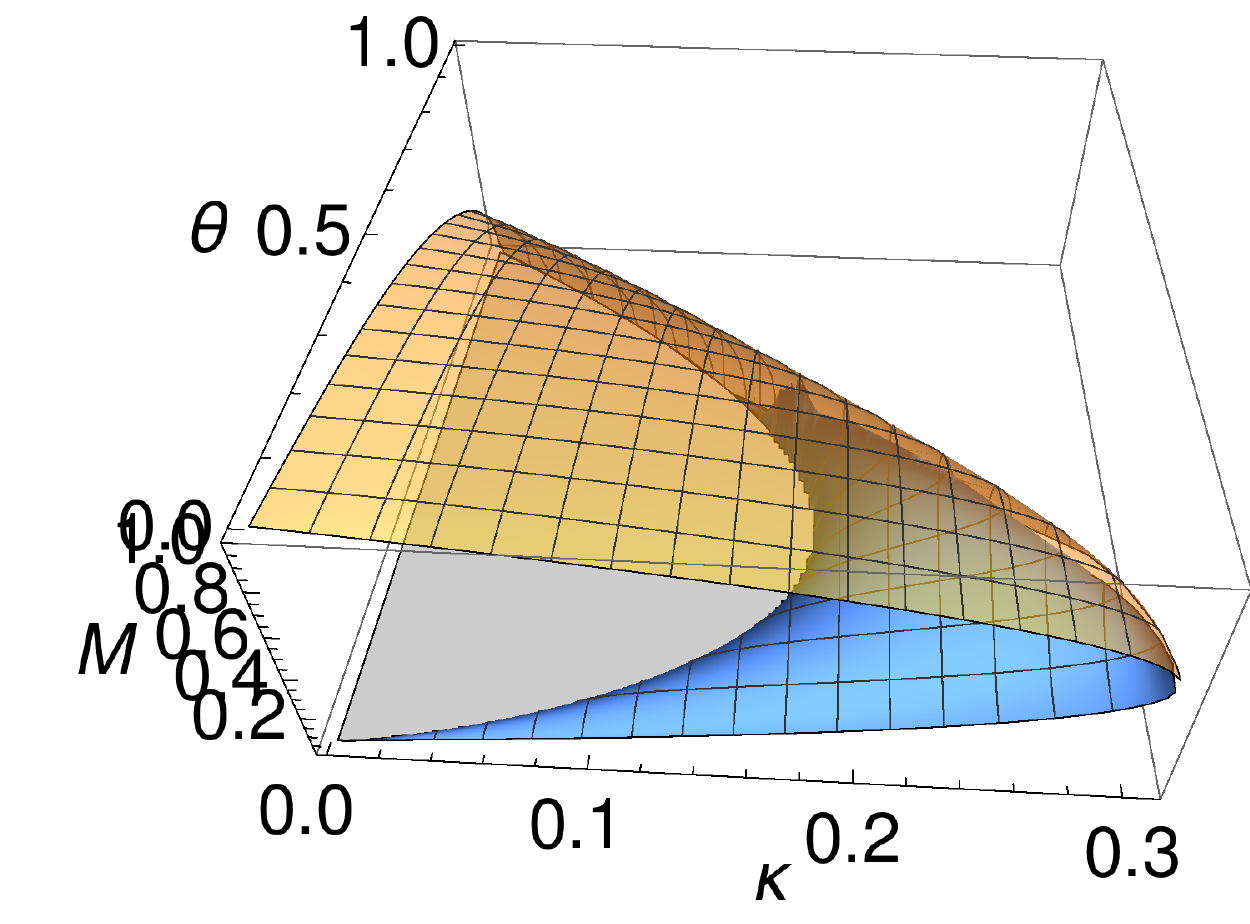}
		\label{4a}}
	\subfigure[]{\includegraphics[width=1.6125in]{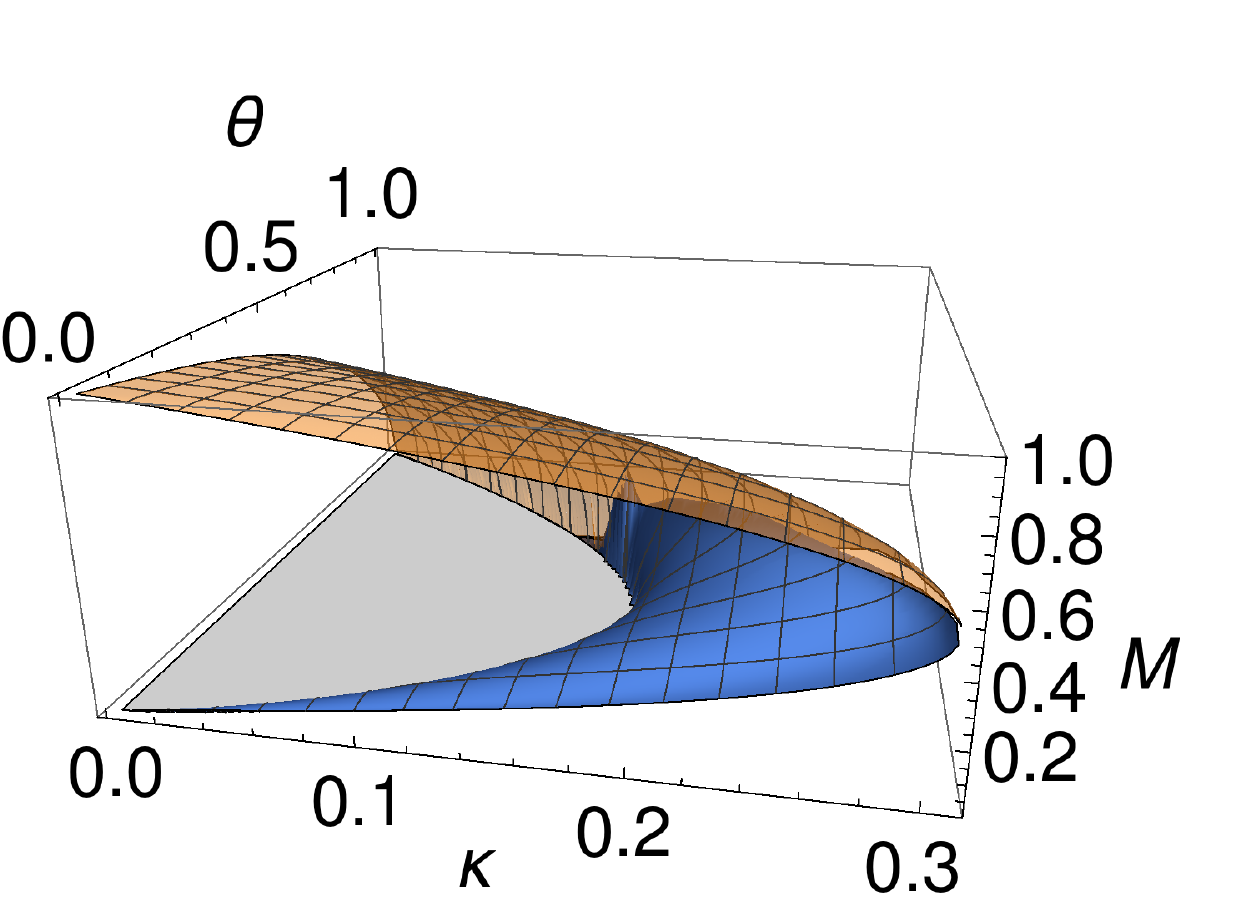}
		\label{4b}}
	\subfigure[$\kappa=0.15$]{\includegraphics[width=1.6125in]{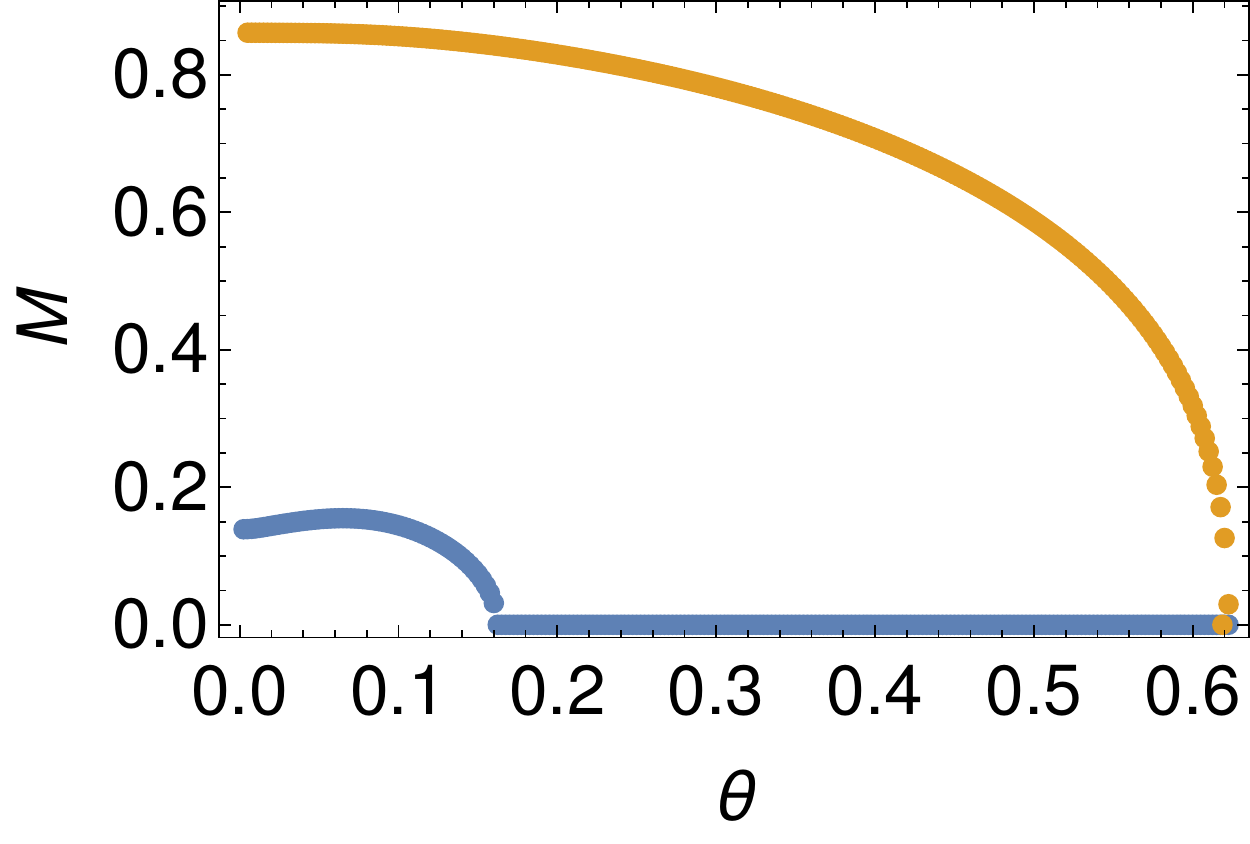}
		\label{4c}}
	\subfigure[$\kappa=0.18$]{\includegraphics[width=1.6125in]{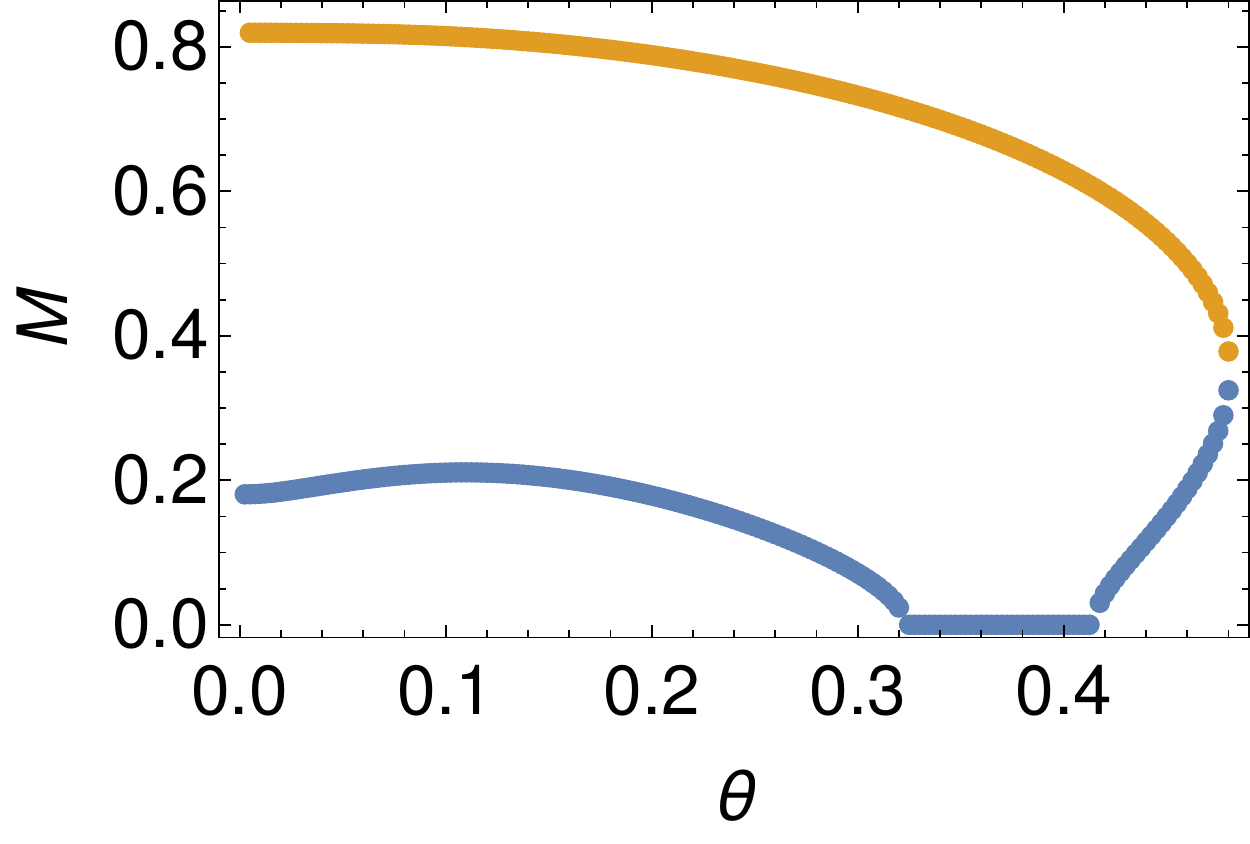}
		\label{4d}}
	\subfigure[$\kappa=0.185$]{\includegraphics[width=1.6125in]{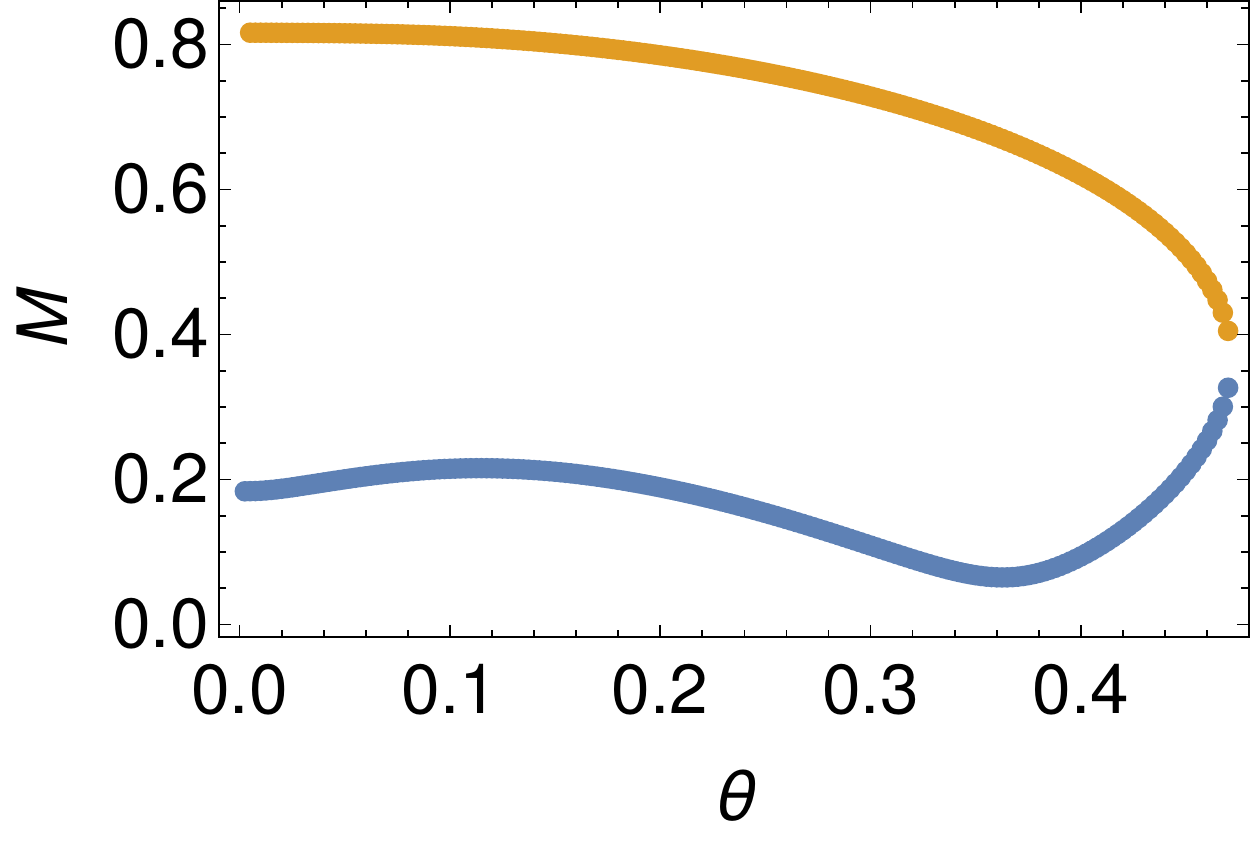}
		\label{4e}}
	\subfigure[$\kappa=0.24$]{\includegraphics[width=1.6125in]{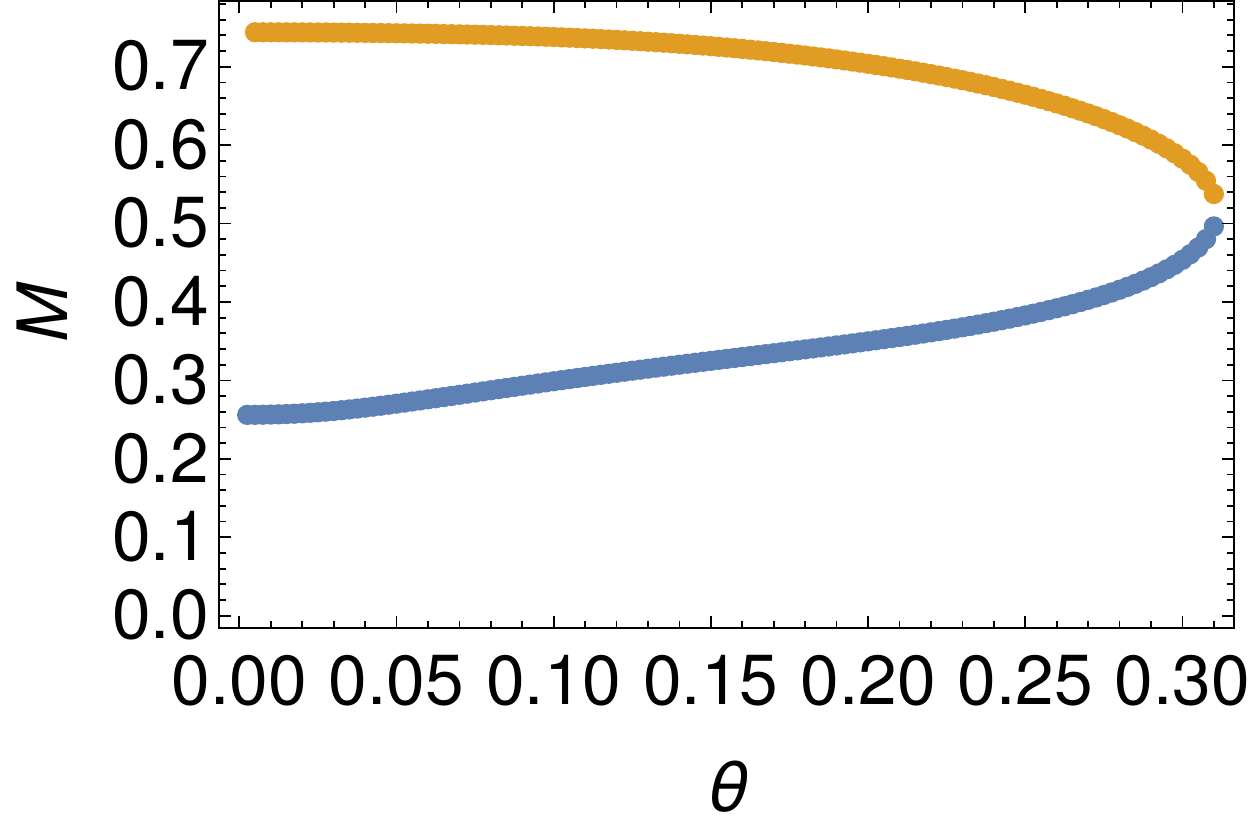}
		\label{4f}}
	\caption{Bifurcation diagrams numerically computed from eq. \eqref{eq:Euler-Lagrange}. Figures \ref{4a} and \ref{4b} show the total magnetization of the buckled solutions versus $\kappa$ and $\theta$. The upper yellow (lower blue) surface stand for the stable (unstable) solution, two different profiles of such solutions at low temperatures can be found in figure \ref{fig:low-temp-profiles}. Panels (c)-(f) are bifurcation diagrams for increasing values of $\kappa$ depicting subcritical and supercritical bifurcations. For $\kappa>\kappa_\nose\sim0.184$, the subcritical bifurcations at the two branches of the bifurcation curve coalesce and an isola stems from the $M=0$ plane. Symmetric results with negative magnetization are omitted for clarity.}
	\label{fig:3d-2d-isola}
\end{figure}

A summary of the above discussion is shown in Table~\ref{tab:phases}. Fig.~\ref{fig:3d-2d-isola}(a) and (b) show the bifurcation diagram of magnetization as a function of $\theta$ and $\kappa$. Panels (c)-(f) depict the magnetization as a function of the temperature for several relevant values of $\kappa$. The bifurcation is always subcritical for the lower branch of the bifurcation curve. As $\kappa>0$ increases, the bifurcation at the upper branch changes from super to subcritical at the tricritical point $\kappa_{c}\simeq 0.159$. The two subcritical bifurcation points merge at the nose point $\kappa_\nose\sim 0.184$. At higher $\kappa$, the buckled phases form an isola separated from the flat configuration. The turning point $\theta_{M}(\kappa)$ at which buckled phases $B-$ and $B$ coalesce marks the boundary between Regions IIIb and I. This is the first-order curve $\kappa_{M}(\theta)$.

\begin{figure}
  \centering
\includegraphics[width=3.in]{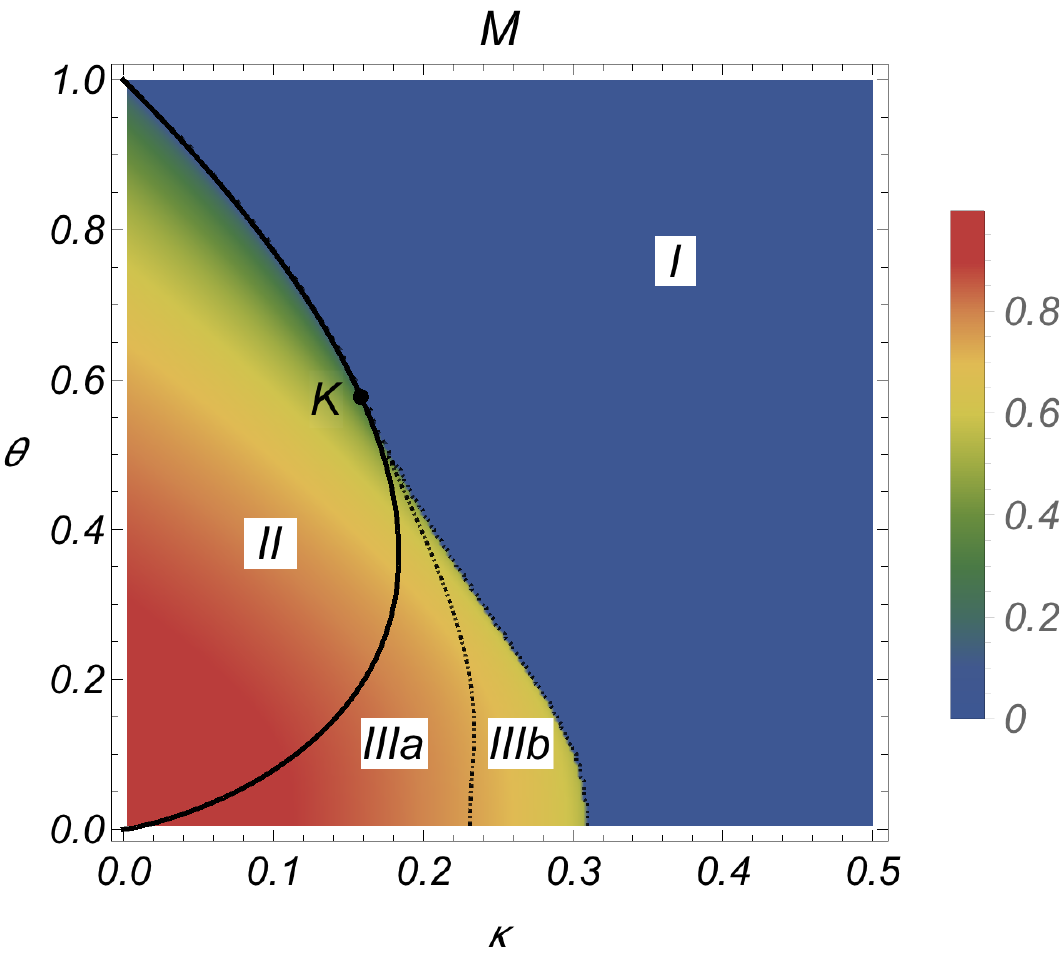}\\
\includegraphics[width=3.in]{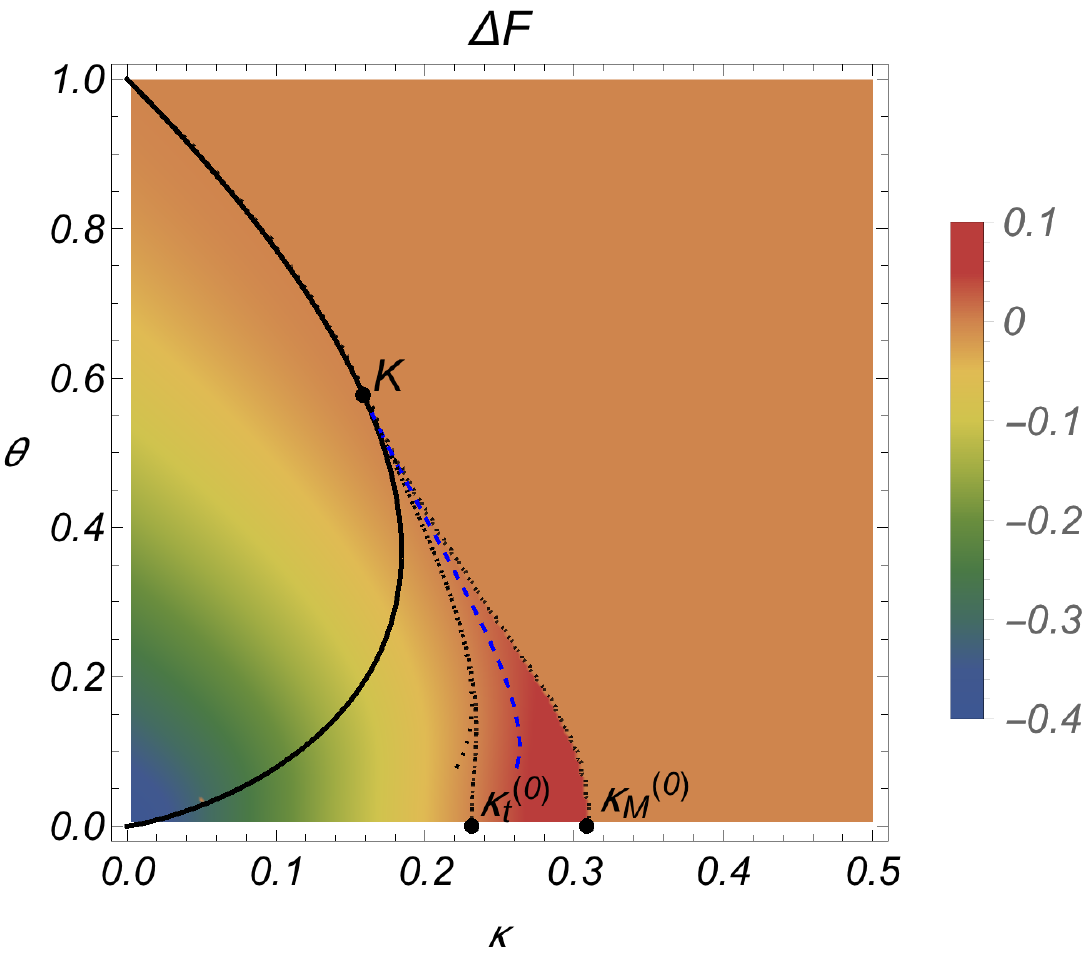}
\caption{ \label{fig:phase-B+}Density plot of the magnetization $M$
  (top panel) and the free energy difference $\Delta F$ (bottom panel)
  over the numerical solution of the Euler-Lagrange equation for the
  phase B+.  Also depicted are (i) the bifurcation line (solid)
  $\kappa_{b}(\theta)$, (ii) the coexistence line (dotted)
  $\kappa_{t}(\theta)$ that separates Regions IIIa and IIIb, at which
  $\Delta F=0$, and (iii) the limit line $\kappa_{M}(\theta)$
  (dotted).  In the free energy panel, we have also plotted the
  analytical expressions close to the critical point for the
  first-order transition lines $\kappa_{t}(\theta)$ and
  $\kappa_{M}(\theta)$, derived in Appendix~\ref{sec:stability}, but
  extended up to low temperatures, namely for $\theta\geq 0.08$.  }
\end{figure}

\begin{figure}
  \centering
\includegraphics[width=3.in]{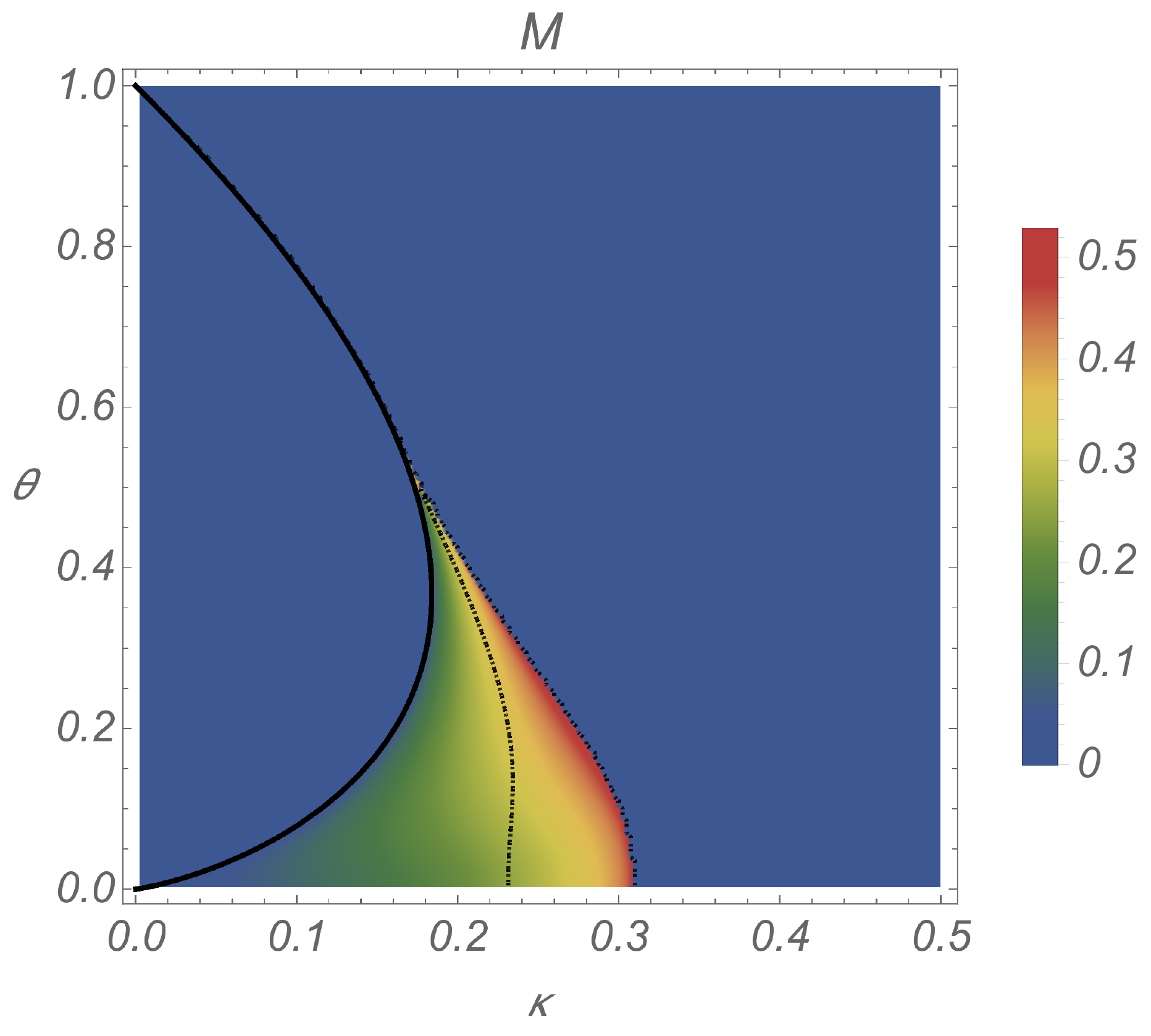}\\
\includegraphics[width=3.in]{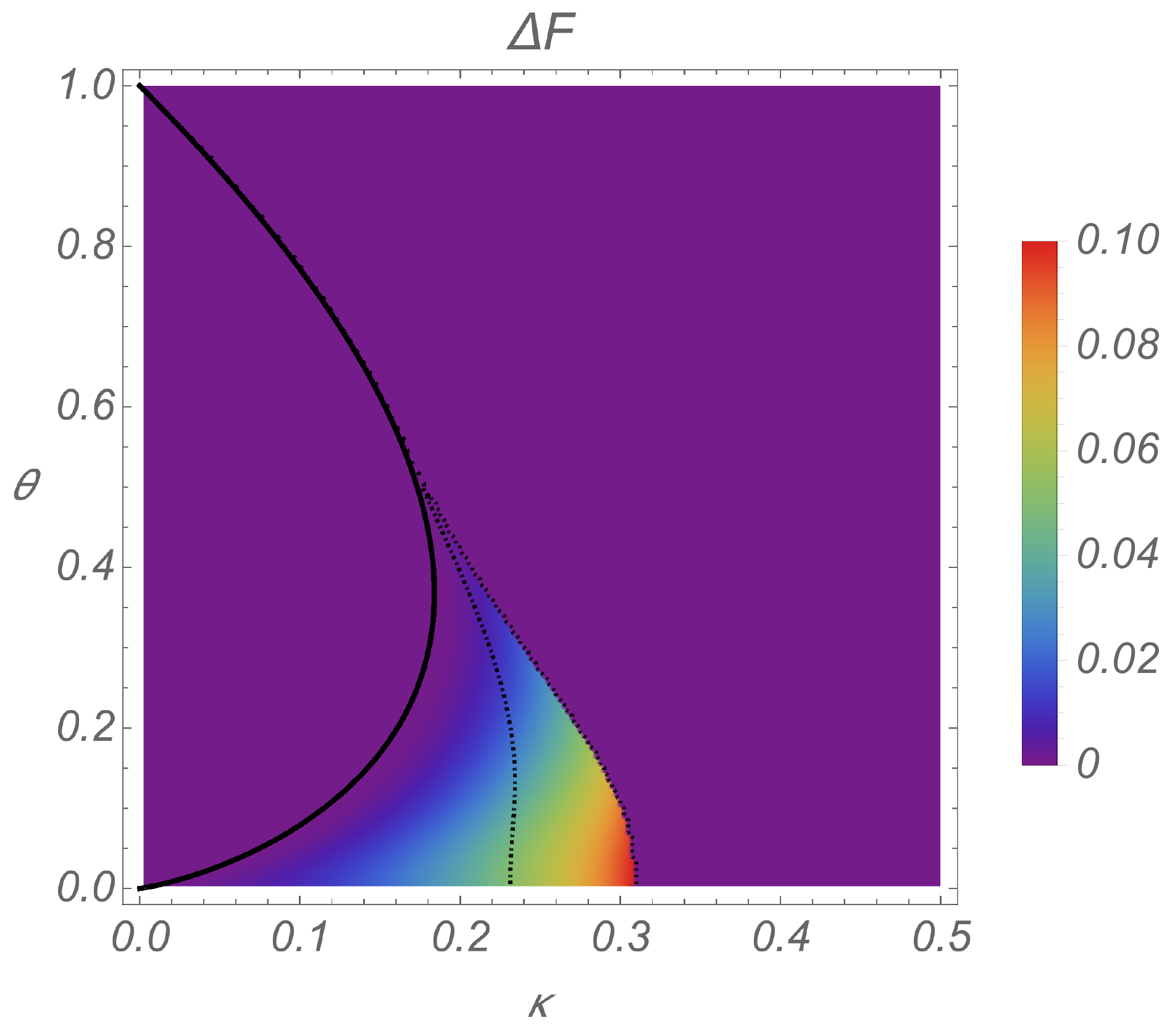}
\caption{\label{fig:phase-B-} Same as Fig.~\ref{fig:phase-B+}, but referred to unstable $B-$ buckled phase.}
\end{figure}

A complementary description to bifurcation diagrams is given in
Figs.~\ref{fig:phase-B+} and \ref{fig:phase-B-}. Figure~\ref{fig:phase-B+}  depicts the
phase diagram of the model showed in Figure~\ref{fig:bifurc}
superimposed on the density plot of phase $B+$ magnetization (top
panel) and free energy (bottom panel). In both panels, it is clearly
observed the change of nature of the transition, from second to
first-order, at the tricritical point $K$. In the bottom panel, the
change of relative stability between phases $B+$ and $L$ at the
coexistence line $\kappa_{t}(\theta)$ is neatly seen, since $\Delta F$
vanishes. Figure~\ref{fig:phase-B-} is completely analogous to
Fig.~\ref{fig:phase-B+}, but for the phase $B-$. Note that
phase $B-$ only exists in region III and is always unstable, $\Delta
F>0$ everywhere.

\section{Bifurcations from the flat string configuration} \label{sec:bifurcation}

In this section, we  calculate the buckled phases that issue from the
flat string near the bifurcation line described in Section
\ref{sec:phase-diagram}. Considerations on the stability of the phases
are included in Appendix \ref{sec:stability}.

\subsection{Pitchfork bifurcations from the flat string configuration }
  Firstly, we expand the free energy about the flat string
  configuration in powers of $u(x)=\epsilon\, U(x)$,
    $\epsilon\ll 1$ and $U=O(1)$, in which $\epsilon$ measures the
amplitude of the string vertical displacement. We define the excess free energy density $\Delta f$
  from the flat configuration (that has energy density $f_L$) as
\begin{eqnarray}
\Delta f(u,u';\kappa,\theta)&\equiv&                        f(u,u';\kappa,\theta)-f_{\fla}(\kappa,\theta), \label{Delta-f-def}
  \\
f_{\fla}(\kappa,\theta)&\equiv&f(0,0;\kappa,\theta)=-\theta
  \ln\left(2\cosh\frac{\kappa}{\theta}\right)\!. \label{eq:f-flat}
\end{eqnarray}
This leads to
\begin{eqnarray}
&&  \Delta f(u,u';\kappa,\theta)=\frac{\epsilon^2}{2\pi^{2}}(U')^{2}+
  \frac{\epsilon^2}{2!}f_{2}(\kappa,\theta)U^{2} \nonumber \\
&& \qquad
+\frac{\epsilon^4}{4!}f_{4}(\kappa,\theta)U^{4}+\frac{\epsilon^6}{6!}f_{6}(\kappa,\theta)U^{6}+O(\epsilon^{8}), \label{eq:f-expansion}
\end{eqnarray}
in which
\begin{equation}\label{eq:fn-def}
\!\!\!f_{n}(\kappa,\theta)\!\equiv\!
\left.\frac{\partial^{n}f(u,u';\kappa,\theta)}{\partial u^{n}}\right|_{u=0}\!=\!-\!\frac{\partial^{n-1}\mu}{\partial u^{n-1}}(0;\kappa,\theta).
\end{equation}
 Here $\mu(u;\kappa,\theta)=-\partial f/\partial u$ is the local magnetization. Using Eq.~\eqref{eq:local-magnetization}, we obtain
\begin{subequations}\label{eq:fn-expressions}
\begin{eqnarray}
f_{2}(\kappa,\theta) & = & - \frac{e^{-2\kappa/\theta}}{\theta}, \\
f_{4}(\kappa,\theta) & = &
                           \frac{e^{-6\kappa/\theta}}{\theta^{3}}\left(3-e^{4\kappa/\theta}\right)\!,\\
f_{6}(\kappa,\theta) & = & -
                           \frac{e^{-10\kappa/\theta}}{\theta^{5}}\left(45-30
                           e^{4\kappa/\theta}+e^{8\kappa/\theta}\right)\!.
\end{eqnarray}
\end{subequations}
 The values of $f_n$ at the bifurcation line \eqref{eq:bifurc-line} are
\begin{subequations}
\begin{eqnarray}\label{eq:fnb}
f_{2,b}&=&-1, \label{eq:f2b} \\
 f_{4,b}&=&\frac{3\theta_{b}^{2}-1}{\theta_{b}^{2}},  \label{eq:f4b}
  \\
  f_{6,b}&=&\frac{-45\theta_{b}^{4}+30\theta_{b}^{2}-1}{\theta_{b}^{4}}. \label{eq:f6b}
\end{eqnarray}
\end{subequations}

Secondly, we expand $\kappa$ and $\theta$ in powers of $\epsilon$:
\begin{subequations}\label{eq:deltas}
\begin{eqnarray}
\delta\kappa(\epsilon)&\equiv&
  \kappa(\epsilon)-\kappa_b=\epsilon^2\kappa_2+\epsilon^{4}\kappa_{4}+O(\epsilon^{6}),\\
\delta\theta(\epsilon)&\equiv&\theta(\epsilon)-\theta_b=\epsilon^2\theta_2+\epsilon^{4}\theta_{4}+O(\epsilon^{6}).
\end{eqnarray}
\end{subequations}
The relation between $\delta\kappa$ and $\delta\theta$
fixes the direction in which we enter the different regions of the
phase diagram. We anticipate that terms containing odd
  powers of $\epsilon$ vanish because $\Delta f$ is invariant under the transformation $U\to -U$.

We now expand $\Delta f$ up to $O(\epsilon^4)$ near the bifurcation line by inserting \eqref{eq:deltas} into \eqref{eq:fn-expressions} and using \eqref{eq:fnb} with
\begin{equation}\label{eq:dfb}
\dfb\equiv f_{2}(\kappa,\theta)-f_{2,b}=\frac{2}{\theta_{b}}\delta\kappa+\frac{1+\ln\theta_{b}}
{\theta_{b}}\delta\theta.
\end{equation}
The result is
\begin{align}
\Delta f\!=\!\frac{\epsilon^2}{2}\!\left(\frac{U'^2}{\pi^2}\!-\!U^2\right)\!\!+\!
\epsilon^{4}\!\left(\!\frac{\varphi_{2}}{2}U^{2}\!+\!\frac{f_{4,b}}{24}U^{4}\!\right)\!\!+\!O(\epsilon^{6}),
  \label{eq:eps-energy}
\end{align}
where
\begin{equation}\label{eq:phi2}
\varphi_{n}=\frac{2\kappa_{n}+\theta_{n}(1+\ln\theta_{b})}
{\theta_{b}}.
\end{equation}
The corresponding Euler-Lagrange equation, to be solved with clamped boundary conditions, is
\begin{eqnarray*}
  \frac{U''}{\pi^2}+U=\epsilon^2\!\left(\varphi_{2}U+\frac{f_{4,b}}{6}U^3\!\right)+O(\epsilon^{4}).
\end{eqnarray*}
 We now insert in this equation the ansatz
\begin{equation}
U(x;\epsilon)=U_0(x)+\epsilon^2 U_2(x)+O(\epsilon^4).\label{eq-U-eps}
\end{equation}
All coefficients of powers of $\epsilon$ are zero separately, which
supplies the hierarchy of equations
\begin{subequations}\label{eq:hierarchy1}
\begin{eqnarray}
\frac{U_0''}{\pi^2}+U_0&=&0, \label{eq:hierarchy1-a}\\
\frac{U_2''}{\pi^2}+U_2&=&                         \varphi_{2}U_0+\frac{f_{4,b}}{6}U_0^3,
\label{eq:hierarchy1-b}
\end{eqnarray}
\end{subequations}
and so on. The boundary conditions are $U_n(0)=U_n(1)=0$. The solution of the first equation is $U_{0}(x)=A \sin\pi x$. Eq.~\eqref{eq:hierarchy1-b} has a
  solution with $U_2(0)=U_2(1)=0$ if its right hand side (rhs) is orthogonal to $\sin\pi x$. This yields the bifurcation equation
\begin{eqnarray}\label{eq:bif1}
\varphi_{2}A+\frac{f_{4,b}}{8}A^3=0,
\end{eqnarray}
Its non-vanishing solutions obey
\begin{equation}
0<A^2=-\frac{8\varphi_{2}}{f_{4,b}}=- 8\theta_b\frac{2\kappa_2+\theta_2
(1+\ln\theta_b)}{3\theta_b^2-1},\label{eq:amp1}
\end{equation}
provided $f_{4,b}\neq 0$ ($\theta_b\neq \theta_c$). In \eqref{eq:amp1} we have substituted $\varphi_{2}$ and $f_{4,b}$ by their explicit expressions.

Let $\kappa$ be the bifurcation parameter, so that $\theta_2=0$. For
$\theta_b>\theta_{c}$, Eq.~\eqref{eq:amp1} produces
$\kappa_2<0$. Then $\kappa<\kappa_b$, and the buckled phases exist only
inside Region II of Figure \ref{fig:bifurc} where the flat string is unstable, i.e., the
bifurcation is supercritical. For $\theta_b<\theta_{c}$
(which also occurs at the whole lower branch of the bifurcation line),
we obtain $\kappa_2>0$, so that $\kappa>\kappa_b$. The buckled phase
bifurcates outside Region II where the flat string is
stable, i.e., the bifurcation is subcritical. Clearly the bifurcating
branches scale as $|\kappa-\kappa_b|^{1/2}$, the usual scaling for a
pitchfork bifurcation.

\subsection{Bifurcation at the tricritical point}

At the tricritical point $K$, the coefficient of $A^3$ in the
bifurcation equation \eqref{eq:bif1} vanishes. We can unfold this
bifurcation by expanding the free energy up to $O(\epsilon^6)$ terms
\cite{Gr74,SyF79} and rescaling the bifurcation
  parameters. If we set $\theta_b=\theta_{c}+\epsilon^2\chi$, with
  $\chi=O(1)$, $f_{4,b}=O(\epsilon^{2})$. Then the leading terms of the coefficients of $U^4$ and $U^6$ in $\Delta f$ are both $O(\epsilon^{6})$. Assuming that $\delta\kappa$ and $\delta\theta$ are
also $O(\epsilon^{4})$ ($\kappa_{2}=\theta_{2}=0$), $\epsilon^{2}\delta f_{2,b}U^{2}=O(\epsilon^{6})$. Then,
\begin{subequations}\label{eq:tricrit_scales}
\begin{eqnarray}
\theta&=&\theta_b+\epsilon^{4}\theta_{4}=\theta_{c}+\epsilon^2\chi+\epsilon^4\theta_4+O(\epsilon^6),\\
\kappa&=&\kappa_{b}+ \epsilon^4\kappa_4+O(\epsilon^6).
\end{eqnarray}\end{subequations}
Keeping terms up to $O(\epsilon^6)$, we obtain
\begin{equation}\label{eq:eps-energy2}
\Delta f \!=\! \frac{\epsilon^2}{2}\!\left[ \frac{(U')^2}{\pi^2}\! -\!U^2
  \right]\!+\! \epsilon^6\!\! \left[ \frac{\varphi_{4,c}}{2}U^2\! +\!\frac{\sqrt{3}\chi}{4} U^4\!+\!\frac{U^6}{20}\right]\!,
\end{equation}
where we have omitted $O(\epsilon^8)$ terms and introduced
  the notation
\begin{equation}\label{eq:phi4c}
\varphi_{4,c}\equiv
\left.\varphi_{4}\right|_{\theta_{b}=\theta_{c}}=\frac{\sqrt{3}}{2}\left[4\kappa_{4}+\theta_{4}(2-\ln
3)\right].
\end{equation}
The corresponding Euler-Lagrange equation is
\begin{align*}
\frac{U''}{\pi^2}+U =  \epsilon^{4}\!\left[\varphi_{4,c}U +\sqrt{3}\chi
U^3+\frac{3}{10}U^5\!\right]+O(\epsilon^{6}),
\end{align*}
to be solved with clamped boundary conditions. We now insert in this equation the ansatz
\begin{equation}
U(x;\epsilon)=U_0(x)+\epsilon^4 U_4(x)+O(\epsilon^{6}),\label{eq-U-eps2}
\end{equation}
thereby obtaining a hierarchy of equations. The equation for $U_0$ is the same as before, whereas $U_4$ solves
\begin{eqnarray} \label{eq:hierarchy2}
\frac{U_{4}''}{\pi^2}+U_{4} =  \varphi_{4,c}U_{0} +\sqrt{3}\chi
U_{0}^{3}+\frac{3}{10}U_{0}^{5}.
\end{eqnarray}
The condition that the rhs of this equation be orthogonal to $\sin\pi
x$ produces the equation for $A$. For $A\neq 0$, it is
\begin{eqnarray}\label{eq:amp3}
A^4+4\sqrt{3} \chi A^2+\frac{8}{\sqrt{3}}\left[4\kappa_4+\theta_4(2-\ln 3)\right] = 0.
\end{eqnarray}
Here we have substituted the explicit expression for $\varphi_{4,c}$.

Let us analyze the solutions of Eq.~\eqref{eq:amp3} for $\theta_4=0$. Then $A$ is a
  function of $\chi$ and $\kappa_{4}$. In Fig.~\ref{fig:bifurc}, the system is just above (below) of the critical point for $\chi>0$ ($\chi<0$) and just outside (inside) the  bifurcation curve for $\kappa_{4}>0$ ($\kappa_{4}<0$). For $\chi>0$, Eq.~\eqref{eq:amp3} has one positive solution $A^{2}>0$ if $\kappa_{4}<0$ ($A^2=0$ for $\kappa_{4}=0$). No real
solutions exist if $\kappa_{4}>0$. For $\chi<0$, Eq.~\eqref{eq:amp3} has one
positive solution $A^{2}>0$ if $\kappa_4<0$ (corresponding to the stable phase $B+$). Depending on the sign of the discriminant of the biquadratic equation, Eq.~\eqref{eq:amp3} has two or zero positive solutions $A^2>0$ for $\kappa_4>0$ (corresponding to stable and unstable phases $B+$ and $B-$). The discriminant of Eq.~\eqref{eq:amp3} vanishes at the curve 
\begin{equation}\label{eq:kappaM}
\kappa_{M}(\theta)=\kappa_{b}(\theta)+\frac{3\sqrt{3}}{8} (\theta-\theta_{c})^{2}.
\end{equation}
Specifically, there are two solutions for $\kappa<\kappa_{M}(\theta)$, denoted by $A_{\pm}^{2}$, $A_{-}^{2}<A_{+}^{2}$, and no
solutions for $\kappa>\kappa_{M}(\theta)$. For $\kappa<\kappa_{M}$,
the solution $A_{-}$ corresponds to phase $B-$ and it issues from the flat configuration as an unstable subcritical
bifurcation at $\theta=\theta_{b}$. The solution $A_{+}$ corresponds to phase $B+$, and it matches at $\theta_{c}$ the only unique phase existing for
$\theta>\theta_{c}$. At the line $\kappa_{M}(\theta)$, phases $B-$ and $B+$ coalesce and disappear, which is consistent with the physical picture of a first-order phase transition. For more details, see
Appendix \ref{sec:stability}.

Note that Eq.~\eqref{eq:amp3} becomes
\begin{equation}
A^2\sim 
-\frac{2}{3\chi}\left[4\kappa_4+\theta_4(2-\ln 3)\right],
\end{equation}
as $\chi\gg 1$. This relation follows from \eqref{eq:amp1} if we
substitute $\theta_{b}=\theta_{c}+\epsilon^2\chi$,
$\theta_{2}=\epsilon^{2}\theta_4$ and
  $\kappa_{2}=\epsilon^{2}\kappa_{4}$ therein. Therefore, as
expected, the bifurcating solution of \eqref{eq:amp3} matches the
solution of the bifurcation equation \eqref{eq:amp1} as
  we move away from the tricritical point.

\subsection{Bifurcation at the turning point}

At the turning point $N$, the coefficient of $A$ in the bifurcation equation \eqref{eq:bif1} becomes $2e\kappa_2$, independent of $\theta_2$. We can unfold this bifurcation by rescaling the bifurcation parameter
\begin{eqnarray}\label{eq:tp_scales}
\kappa=\underbrace{\frac{1}{2e}}_{\kappa_{\nose}}+ \epsilon^4\kappa_4+O(\epsilon^6),
\end{eqnarray}
and expanding the coefficient of $U^2$ in the free energy up to $O(\epsilon^4\theta_2^2)$ terms. Inserting the result in Eq.~\eqref{eq:eps-energy}, we obtain
\begin{eqnarray}
\Delta f&=&\frac{\epsilon^2}{2}\!\left[ \frac{(U')^2}{\pi^2} -U^2 \right]\!\nonumber\\
&&+\epsilon^4\!\left[\!\left(\kappa_4+\frac{e}{4}\theta_2^2\right)\!e U^2\!+\frac{3-e^2}{24} U^4\right]\!+O(\epsilon^6). \label{eq:eps-energy3}
\end{eqnarray}
The corresponding Euler-Lagrange equations are
\begin{eqnarray*}
\frac{U''}{\pi^2}+U=\epsilon^2\!\left[2e\!\left(\kappa_4+\frac{e}{4}\theta_2^2\right)\! U-\frac{e^2-3}{6} U^3\!\right]\!+O(\epsilon^4),
\end{eqnarray*}
to be solved with clamped boundary conditions. Inserting
\eqref{eq-U-eps} into this formula and equating like powers of
$\epsilon$, we obtain a hierarchy of equations.

Again, the solution of the first equation of the hierarchy with
clamped boundary conditions is $U_{0}(x)=A\sin\pi x$. The second
equation is
\begin{equation*}
\frac{U''_2}{\pi^2}+U_2=2e\!\left(\kappa_4+\frac{e}{4}\theta_2^2\right)\!U_{0}-\frac{e^2-3}{6} U_{0}^{3}.
\end{equation*}
 This equation has a solution that satisfies clamped boundary
 conditions provided its rhs is orthogonal to $\sin\pi x$, which yields
 \begin{equation*}
2e\!\left(\kappa_4+\frac{e}{4}\theta_2^2\right)\! A-\frac{e^2-3}{8} A^3=0.
\end{equation*}
The nontrivial solution of this equation satisfies
\begin{equation}\label{eq:amp4}
0<A^2=\frac{4e(4\kappa_4+e\theta_2^2)}{e^2-3}.
\end{equation}
Note that $\kappa_{4}=-e\theta_{2}^{2}/4$ is nothing but the
  lowest approximation to the bifurcation curve in the vicinity of the
  nose, written in the scaled variables.

 Equation~\eqref{eq:amp4} implies that buckled solutions
  stem continuously from the parabola $\kappa_4=-e\theta_2^2/4$ and
  exist outside it.  These buckled states (corresponding to phase $B-$) bifurcate subcritically
  at $\theta_2^{(1,2)}=\pm 2\sqrt{-\kappa_{4}/e}$. The corresponding temperatures are on the
  upper and lower branches of the bifurcation curve, respectively. At the turning point
  $\kappa_{4}=0$ and the two bifurcation points merge. For
  $\kappa_{4}>0$ ($\kappa>\kappa_{\nose}$), there is a single unstable
  buckled state given by \eqref{eq:amp4}, for points close enough to
  the bifurcation curve.

  Note that the stable phase $B+$ cannot be predicted by the
  bifurcation analysis near the nose, since the corresponding string
  profile is not close to the flat solution therein. We know that, for
  fixed $\kappa>\kappa_{\nose}$, both buckled phases $B\pm$ coalesce
  at the boundary between Regions IIIb and I in Figure
  \ref{fig:phase-B+}. These buckled string configurations persist as
  the temperature $\theta\to 0+$ for all spin-spin couplings
  $\kappa<\kappa_{M}^{(0)}=\pi^2/32$, as indicated in Section
  \ref{sec:low-temp-profile}.

\section{Low temperature behavior}
\label{sec:exact-sols}

\subsection{Low temperature profiles: Exact solution of the Euler-Lagrange equation}
\label{sec:low-temp-profile}

At very low temperatures, such that $\exp(-2\kappa/\theta)/\theta<1$
in Fig.~\ref{fig:bifurc}, there are buckled solutions in
  addition to the stable flat profile. We calculate exactly their profiles below.

In fact, for $\theta\ll |u|$, the local magnetization $\mu$ of Eq.{\eqref{eq:local-magnetization} reduces to
\begin{equation}\label{eq:magnet-theta-0}
\mu(u;\kappa,\theta=0^{+})=\sgn(u) \eta(|u|-2\kappa),
\end{equation}
where $\eta(x)$ is the Heaviside step function.  Substituting
Eq.~\eqref{eq:magnet-theta-0} into \eqref{eq:Euler-Lagrange}, we find
$u''=0$ if $|u|<u_{0}=2\kappa$, and $u''=\pm\pi^{2}$ if
$|u|>2\kappa$. Then $u(x)$ is a linear function if $|u|<2\kappa$, and
a parabola if $|u|>2\kappa$.  Due to the clamped boundary conditions,
buckled solutions with a single extremum (no internal nodes) are
linear close to the boundaries, $x\in(0,x_{0})$ or $x\in(1-x_{0},1)$,
and have a parabolic profile in the bulk $x\in(x_{0},1-x_{0})$. The
condition $|u(x_0)|=2\kappa$ produces the condition
$\pi^2 x_0 (1-2x_0) = 4\kappa$ whose solutions $x_{0,1}$ and $x_{0,2}$
are
\begin{equation}\label{eq:x0-values}
x_{0,j}=\frac{1}{4}\left(1+(-1)^j\sqrt{1-\frac{\kappa}{\kappa_{M}^{(0)}}}\right)\!,\quad j=1,2,
\end{equation}
for $\kappa<\kappa_M^{(0)}=\pi^2/32$. We have $x_{0,1}<1/4<x_{0,2}$, $x_{0,1}+x_{0,2}=1/2$. If $\kappa>\kappa_{M}^{(0)}$, these rippled low-temperature profiles are not possible and the only solution is $u=0$.

\begin{figure}
  \centering
  \includegraphics[width=3.25in]{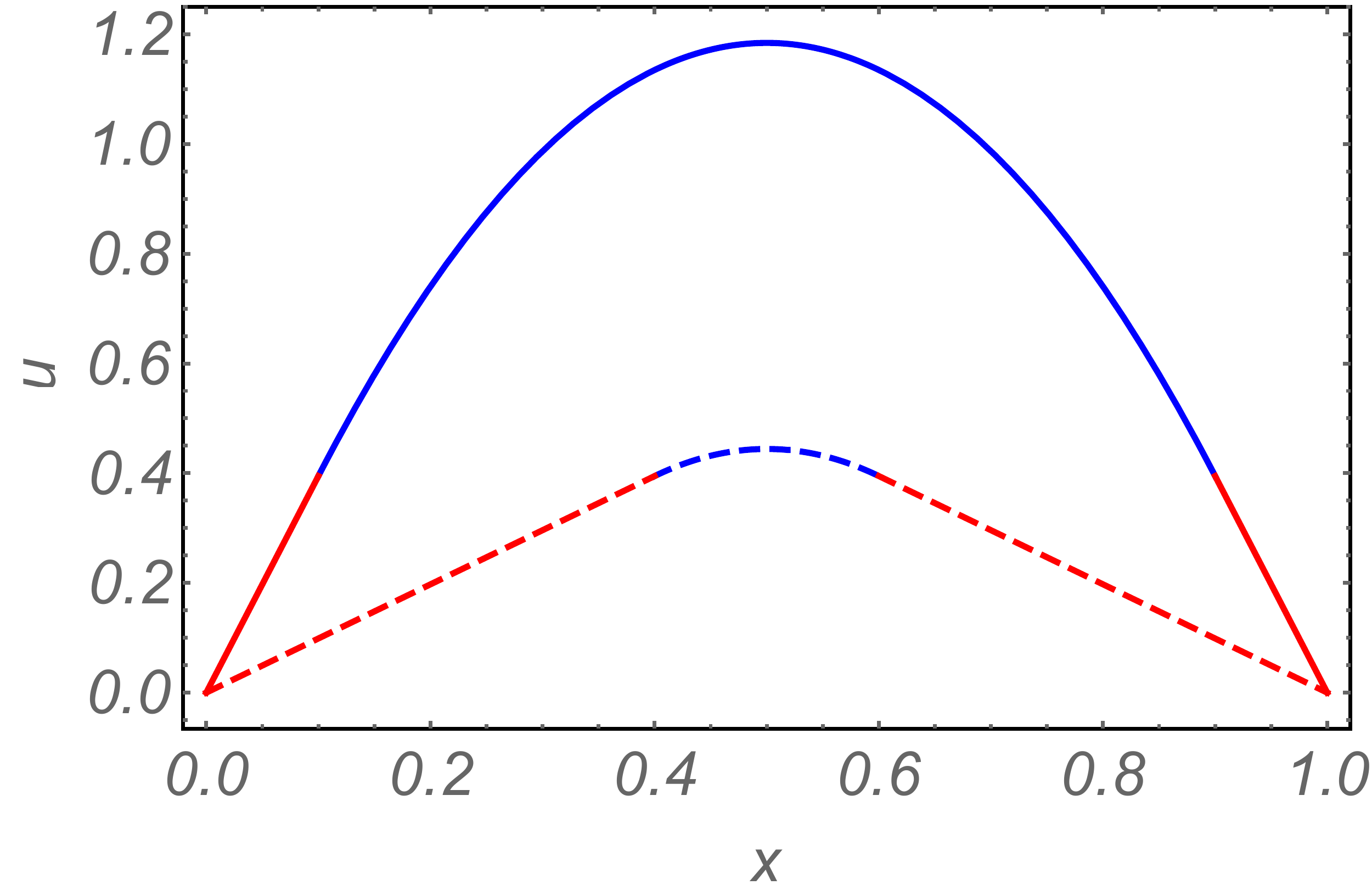}
  \caption{Low temperature non-flat string profiles. They comprise two
    linear zones of width $x_0$ near the endpoints of the chain (in
    red) and a parabolic zone in the middle of the system of width
    $1-2x_0$ (in blue). Spins are ordered
    antiferromagnetically in the linear zones and ferromagnetically in the central parabolic zone, see
    Fig.~\ref{fig:esq_na}. In this plot,
    $\kappa=\pi^{2}/50<\kappa_{t}^{(0)}$, which gives two possible
    values of $x_{0}$: $x_{0,1}=1/10$ (solid line) and $x_{0,2}=2/5$
    (dashed line). The buckled profile corresponding to $x_{0,2}$ is always unstable. The string profile
    corresponding to $x_{0,1}<1/8$ gives the absolute minimum of
    the free energy, and the flat string is
    metastable.  }
  \label{fig:low-temp-profiles}
\end{figure}

Fig.~\ref{fig:low-temp-profiles} shows two of these profiles for an
appropriate value of $\kappa$. The same functions multiplied by -1 are
also stationary solutions. In these string profiles, the pseudo-spins
exhibit antiferromagnetic order close to the boundaries and
ferromagnetic order in the bulk, see below.

 The profiles with $x_{0,1}<1/4$ produce a relative
  minimum of the free energy and are stable whereas those with
  $x_{0,2}>1/4$ are unstable \cite{note1}, as proven in Appendix
  \ref{sec:stability-low-temp-prof}. Thus, for
  $\kappa<\kappa_{M}^{(0)}$, the buckled profiles with $x_{0,1}<1/4$
and the flat string are stable and the unstable profiles with
$x_{0,2}>1/4$ separate them. The stable and unstable buckled profiles
coalesce and disappear at $\kappa=\kappa_{M}^{(0)}$
($x_{0,1}=x_{0,2}=1/4$). This allows us to identify the
  buckled profiles with $x_{0,1}$ and $x_{0,2}$ as the low temperature
  limits of phases $B+$ and $B-$, respectively.

By direct integration, we can show that the absolute minimum of the
free energy corresponds to the buckled configurations with $x_{0,1}$
if $0<\kappa<\kappa_{t}^{(0)}=3\pi^{2}/128$ ($0< x_{0,1}<x_{t}=1/8$).
For $\kappa_{t}^{(0)}<\kappa<\kappa_{M}^{(0)}$ ($x_{t}<x_{0,1}<1/4$),
the free energy of the flat string is smaller than that of the buckled
configurations with $x_0=x_{0,1}$. Thus the flat string profile is
metastable for $0<\kappa<\kappa_{t}^{(0)}$ and stable for
$\kappa_{t}^{(0)}<\kappa<\kappa_{M}^{(0)}$. The situation is reversed
for the buckled configurations with $x_0=x_{0,1}$.  At
$\kappa=\kappa_{t}^{(0)}$ there is a first order phase transition,
where the buckled phase with $x_{01}$ and the flat string coexist.
Consistently, the first-order derivatives of the free energy change
discontinuously at $\kappa=\kappa_{t}^{(0)}$. In fact, as $\kappa$
increases past ${\kappa_{t}^{(0)}}$, $M$ and $\mathcal{DL}$ jump from
$M=3/4$ and $\mathcal{DL}=3/4$ (buckled phase with $x_{0,1}$) to
$M=0$ and $\mathcal{DL}=1/2$ (flat phase).

\subsection{Spin configurations of the low temperature buckled string
  states}

What are the spin configurations at buckled string states? It turns out that the spins form antiferromagnetic domains near the boundaries and ferromagnetic domains in the central region of the string. To see this, we derive their marginal probability
 $\mathcal{P}(\bm{\sigma})$ by integrating the canonical distribution
$\exp(-\calH/T)$ over the string degrees of freedom. The result is
\begin{equation}\label{eq:Prob-spins}
 \mathcal{P}(\bm{\sigma})\! \propto\! e^{-\mathcal{H}_{\text{eff}}(\bm{\sigma})/\theta}\!,
 \, \mathcal{H}_{\text{eff}}(\bm{\sigma})\!=\! \kappa \bm{\sigma^T\!
      J\sigma}\!-\!\frac{\pi^{2}}{2N^{2}}\bm{\sigma^T \Lambda\sigma}.
\end{equation}
Here, the effective spin Hamiltonian $\mathcal{H}_{\text{eff}}$ contains a nearest neighbor antiferromagnetic interaction given by
\begin{equation}\label{eq:J-matrix}
\bm{J}_{ij}=\frac{1}{2}\left(\delta_{i,j+1}+\delta_{i,j-1}\right)
\end{equation}
and a long-ranged ferromagnetic interaction given by
\begin{equation}\label{eq:Lambda}
\bm{\Lambda}_{ij}=\frac{1}{N+1} j (N-i+1) > 0, \ \  \forall i\ge j,
\quad \bm{\Lambda}_{ij}=\bm{\Lambda}_{ji},
\end{equation}
 which is derived in
  Appendix~\ref{sec:low-temp-spins}. Phase transitions in a one-dimensional model stem
  from this effective long range interaction, similarly to the situation found in other spin-oscillator
  models \cite{pre12bon,jstat10,jstat10a}.

We focus on the low temperature limit as $\theta\to 0^{+}$: therein,
the equilibrium probability concentrates in the spin
configuration that corresponds to the absolute minimum of
$H_{\text{eff}}$. The long-range ferromagnetic
  interaction \eqref{eq:Lambda} is stronger for the
  pseudo-spins located near the center of the system than for those
  close to the boundaries. Therefore, as the intensity of the
antiferromagnetic interaction $\kappa$ increases, the absolute
minimum of $H_{\text{eff}}$ moves from the completely ferromagnetic
configuration to one that is antiferromagnetic at the boundaries and
ferromagnetic in the bulk. See Appendix \ref{effective-H} for
details.

In light of the previous discussion, we restrict ourselves to states that are antiferromagnetic at
the boundaries and ferromagnetic in the center. Note that this restriction includes
completely antiferromagnetic and ferromagnetic states. We label the states by the number $n_{a}=1,3,5,\ldots, N/2$ of spins at the antiferromagnetic boundary regions; see Fig.~\ref{fig:esq_na}. Moreover, we
denote by $\mathcal{H}_{\text{eff}}(n_a)$ the effective potential for
such a configuration. In Appendix~\ref{effective-H}, we find
\begin{eqnarray}
  \mathcal{H}_{\text{eff}}(n_a)=(n_{a}-1)\Bigl\{ \frac{\pi^2}{6N^2}
                                                    [&& N (3 + n_a )3-21
                                                    - 13 n_a
                                                    \nonumber \\
&&- 4 n_a^2 ]- 4\kappa\Bigr\}.
   \label{eq:Hna}
\end{eqnarray}
The origin of energy is such that $\mathcal{H}_{\text{eff}}(n_a=1)=0$.

\begin{figure}
	\centering
	\includegraphics[width=3.25in,height=2.1in]{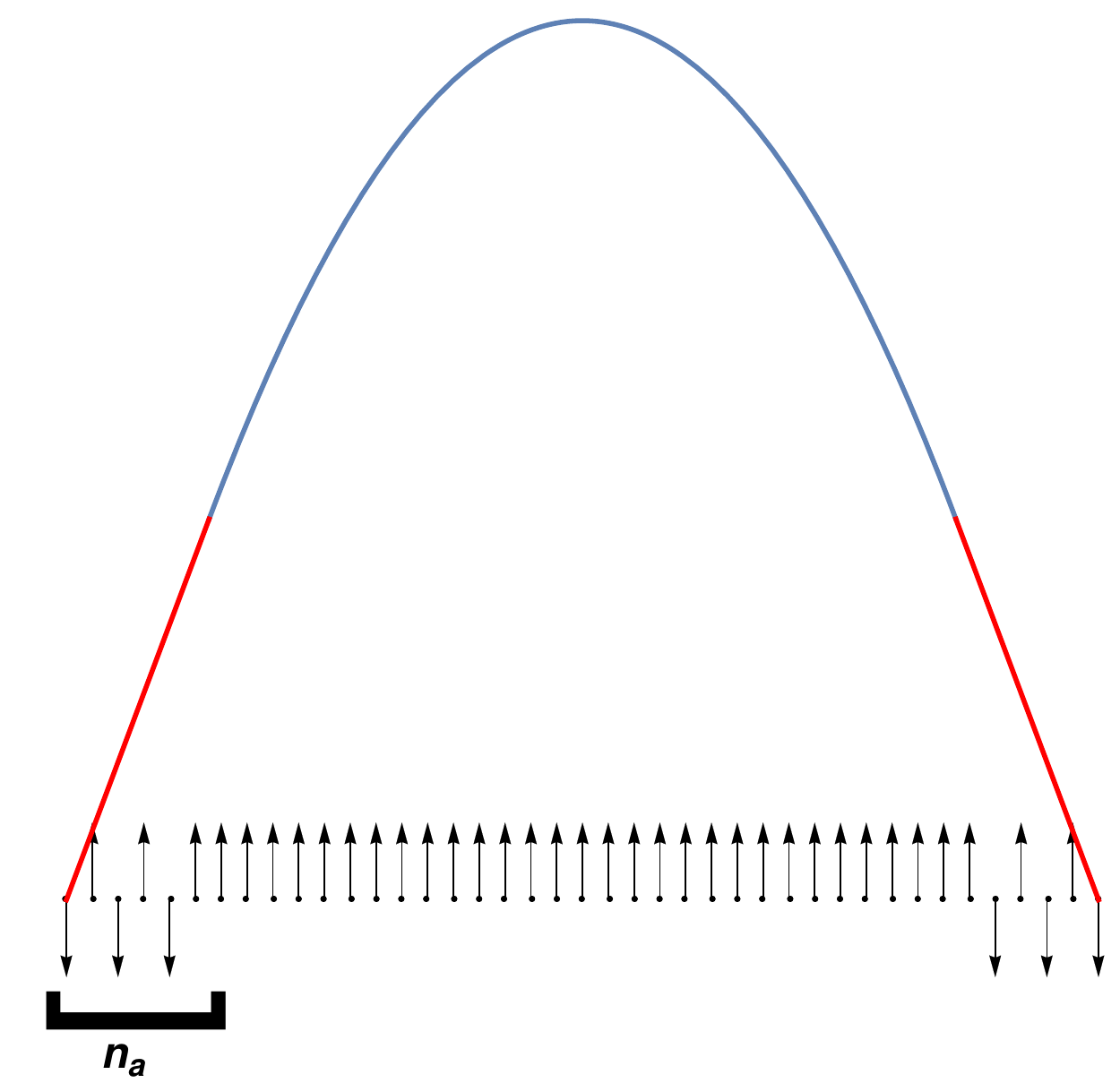}
	\caption{\label{fig:esq_na} Qualitative graph for the typical
          low-temperature configurations for the pseudo-spins and the
          string. In the antiferromagnetic regions close to the
          boundaries, there is no net magnetization and thus the
          string has a linear profile ($u''=0$, red). In the
          ferromagnetic region in the bulk, the string takes a
          parabolic shape ($u''=-1$, blue). We are plotting a system
          with $N=41$ pseudo-spins and $n_{a}=5$, which is the number
          of antiferromagnetic links at either boundary.  }
\end{figure}

Depending on the value of $\kappa$, $\mathcal{H}_{\text{eff}}(n_a)$
has one or two minima, as seen in Fig.~\ref{fig:H-eff-vs-na}.  For
$\kappa=0$, the completely ferromagnetic configuration gives the
minimum of $H_{\text{eff}}$, as expected on physical grounds. On the
other hand, as $\kappa$ increases, there appear several relevant
values of $\kappa$, namely
\begin{subequations}\label{eq:kappas-spins}
\begin{eqnarray}
\kappa_0&=&\frac{\pi^{2}}{4}\frac{N - 1}{N(N + 1)}, \\
\kappa_1&=&\frac{\pi^{2}}{384}\frac{9 N^2+6N-47}{N^{2}}, \\
\kappa_2&=&\frac{\pi^{2}}{96} \frac{ 3 N^2+ 6 N -5 }{N^{2}},
\end{eqnarray}
\end{subequations}
the physical meaning of which are discussed below. First, for
$\kappa=\kappa_{0}$, the configurations with $n_{a}=1$ and $n_{a}=0$
share the same value of $\mathcal{H}_{\text{eff}}$. This marks the
onset of the antiferromagnetic ordering at the boundaries, although
for a large system this ordering is only relevant when $n_{a}/N$
becomes of the order of unity. In fact, for large $N$,
  $\kappa_{0}$ is proportional to $N^{-1}$ , whereas both $\kappa_{1}$
  and $\kappa_{2}$ become independent of $N$.  Second, at $\kappa_1$,
the relative minimum of $\mathcal{H}_{\text{eff}}$ has the same value
as the completely antiferromagnetic configuration. Finally, at
$\kappa_2$, this relative minimum disappears and the only stable
configuration is that of the absolute minimum for $n_{a}=N/2$, that
is, the completely antiferromagnetic configuration.

\begin{figure}
\centering
\subfigure[$\kappa=\kappa_0$]{\includegraphics[width=1.6125in]{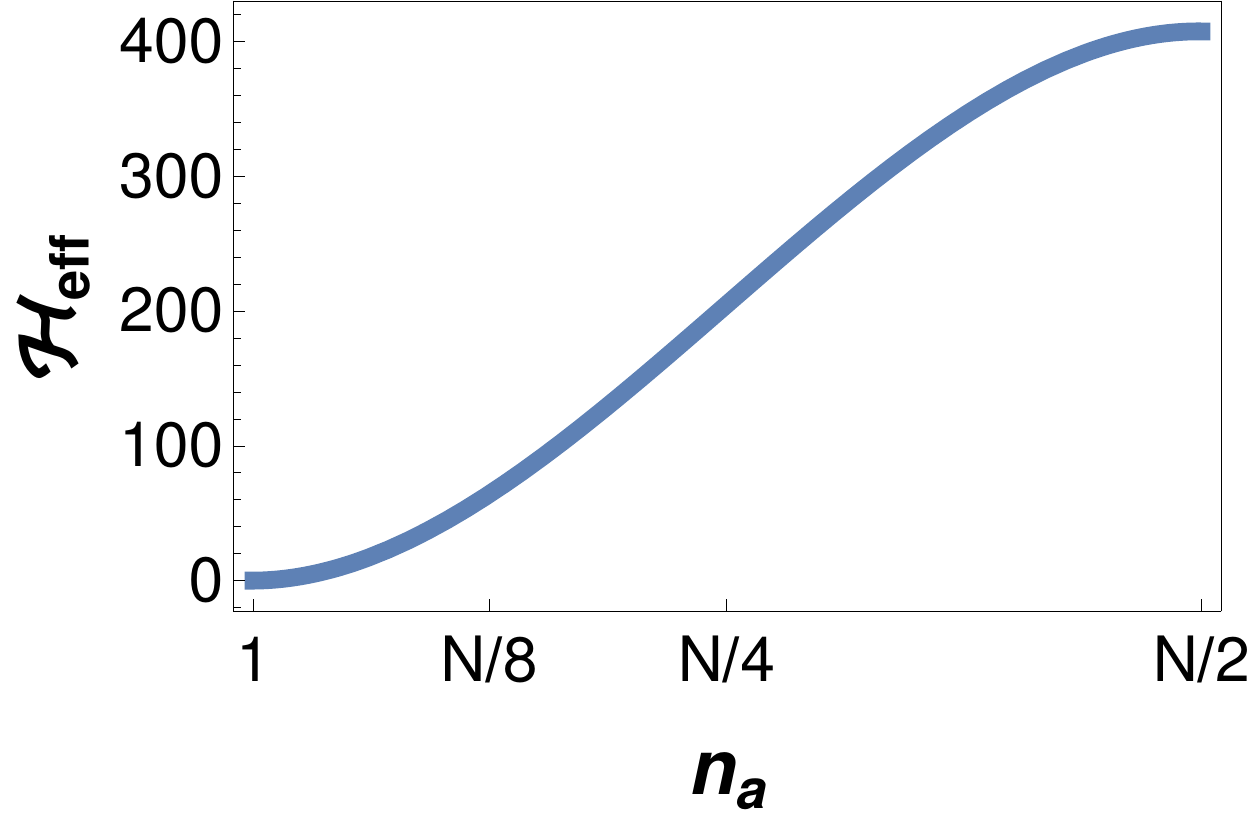}
\label{a}}
\subfigure[$\kappa=\kappa_0+\frac{3}{5}(\kappa_2-\kappa_0)$]{\includegraphics[width=1.6125in]{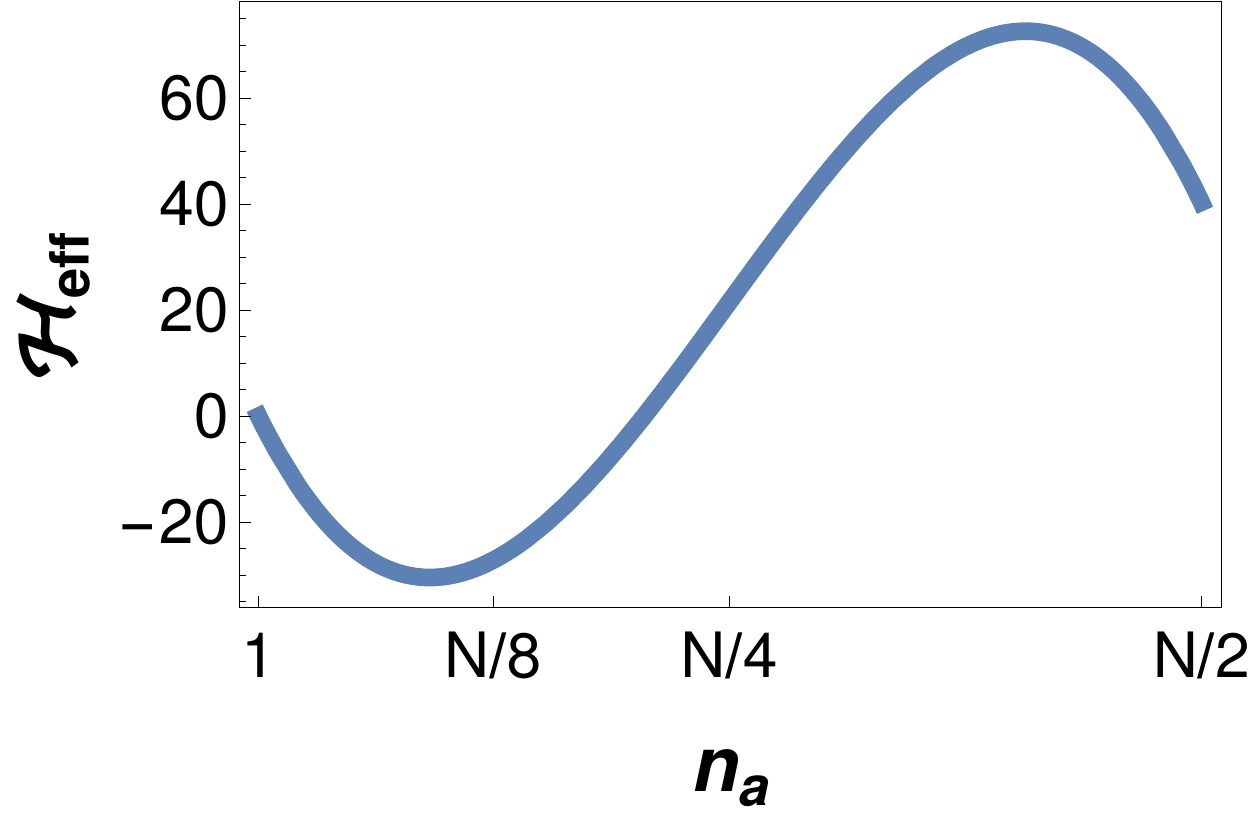}
\label{d}}
\subfigure[$\kappa=\kappa_1$]{\includegraphics[width=1.6125in]{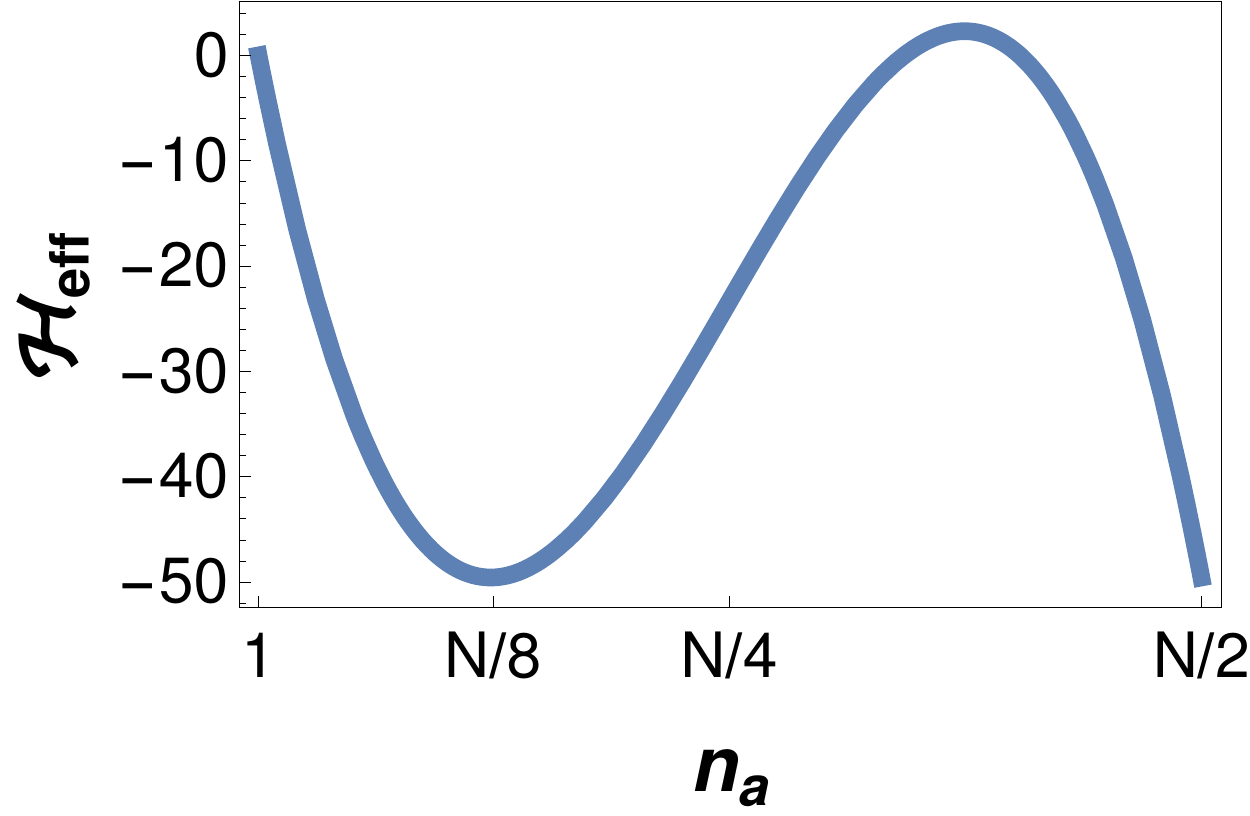}
\label{e}}
\subfigure[$\kappa=\kappa_0+\frac{4}{5}(\kappa_2-\kappa_0)$]{\includegraphics[width=1.6125in]{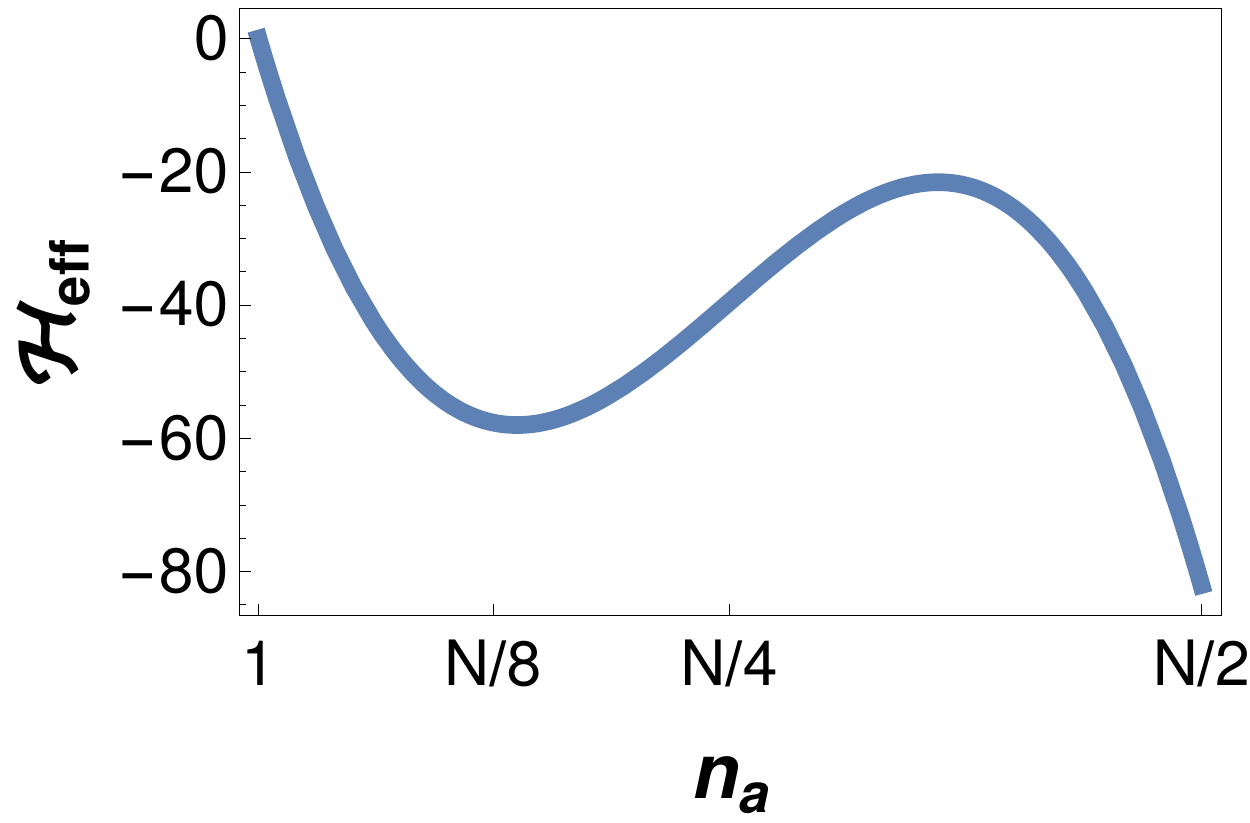}
\label{f}}
\subfigure[$\kappa=\kappa_2$]{\includegraphics[width=1.6125in]{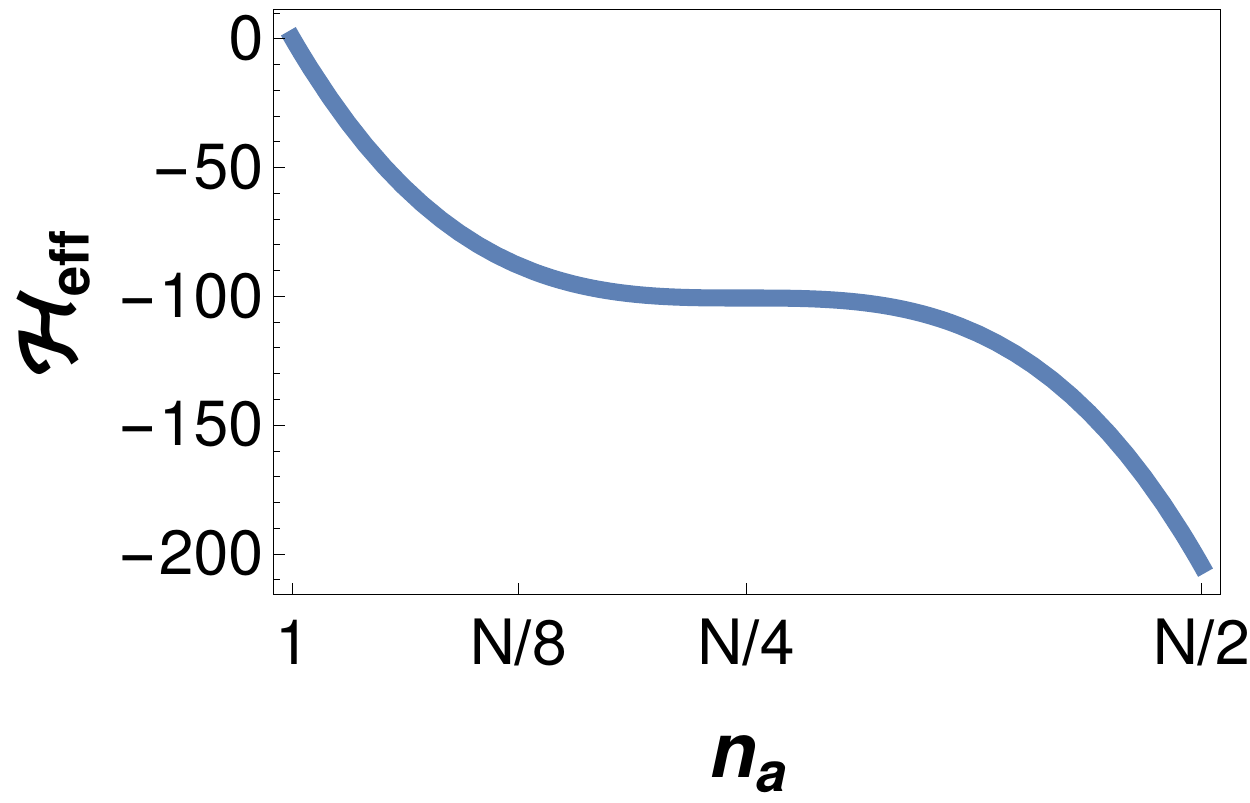}
\label{g}}
\caption{\label{fig:H-eff-vs-na} $\mathcal{H}_{\text{eff}}$ (in arbitrary units) as a
  function of the number of antiferro links ${n_a}$, for different
  values of $\kappa$.  In the limit as $\theta\to 0^{+}$, the
  probability concentrates in the \textit{absolute} minimum of
  $H_{\text{eff}}$. (a) $\kappa=\kappa_{0}$, the absolute minimum
   occurs at $n_{a}=1$. (b) $\kappa_{0}<\kappa<\kappa_{1}$: the
  absolute minimum shifts to greater
  values of $n_a$, and a new minimum appears at the completely
  antiferromagnetic configuration $n_{a}=N/2$. The
    minima correspond to phases $B+$ and $L$ and the maximum
   between them to phase $B-$. (c) $\kappa=\kappa_{1}$, $\mathcal{H}_{\text{eff}}$ is the same at the
  relative minimum and at the completely antiferromagnetic state with $n_{a}=N/2$. (d) $\kappa_{1}<\kappa<\kappa_{2}$, $H_{\text{eff}}$ is lowest for the completely antiferromagnetic
  configuration. (e) $\kappa=\kappa_{2}$, the relative
  minimum disappears and the only equilibrium state is the completely
  antiferromagnetic one. }
\label{fig:3}
\end{figure}

The situation described above is illustrated in
Fig.~\ref{fig:H-eff-vs-na}, in which we plot
$\mathcal{H}_{\text{eff}}$ as a function of $n_{a}$, for different
values of $\kappa$.  Of course, in the large $N$ limit, the values of
$\kappa$ at which there are changes in the stability of the solution
are in perfect agreement with those obtained from the analysis of the
solution of the Euler-Lagrange equation \eqref{eq:Euler-Lagrange} for
the string profile: $\kappa_{1}$ and $\kappa_{2}$ tend to
$\kappa_{t}^{(0)}$ and $\kappa_{M}^{(0)}$, respectively. The
completely antiferromagnetic configuration leads to an almost flat,
wrinkled, string whereas the completely ferromagnetic distribution
corresponds to a buckled configuration, with a definite sign of the
curvature. Accordingly, the low temperature phase, comprising
antiferromagnetic boundaries and a ferromagnetic bulk yields a buckled
string with linear ($u"=0$) boundaries, as depicted in
Figs.~\ref{fig:low-temp-profiles} and \ref{fig:esq_na}.

\section{Conclusions}\label{sec:conclusions}

Despite its simplicity, the $1$d string model contains the key
ingredients that lead to the emergence of wrinkled and buckled phases
in graphene.  The transversal displacements $u_{i}$ are coupled to internal degrees of freedom, modeled by spin variables $\sigma_{i}$. The latter have two competing
interactions: (i) an on-site interaction with their corresponding
displacements, and (ii) an antiferromagnetic interaction (of strength $\kappa$) between nearest
neighbor spins.

A coarse-grained approach, where internal degrees of freedom are integrated out to give rise to an effective free energy for the string deformation, entails that the string curvature is controlled
  by the local magnetization and the flat string phase $L$ becomes
  unstable inside a bifurcation line $\kappa_{b}(\theta)$ whose inverse function is two-valued. For a given $\kappa$, lowering the temperature $\theta$ produces
buckled string profiles with
non-zero global magnetization. For low enough temperatures,
the short-ranged antiferromagnetic interaction: (i) modifies the
buckled profiles, introducing an antiferromagnetic region close to the
boundaries, and (ii) makes the flat string metastable.

 Fig.~\ref{fig:bifurc}
  and Table~\ref{tab:phases} provide a summary of the different phases, their domains of
  existence and their stability. In region I, the antiferromagnetic interaction prevails and only the flat phase $L$ exists. In region II, the long-range ferromagnetic interaction
  dominates and there appears a stable buckled phase $B+$. For each $\theta$ in Region III, there is a competition between the
  ferromagnetic interaction that induces global buckling and the
  antiferromagnetic interaction that favors the flat phase. Therein,
  both the flat phase $L$ and the buckled phase $B+$ are locally
  stable minima of the free energy. In addition, there appears
  an unstable buckled phase $B-$ that separates these minima.

A key element in the observed behavior is the existence of a
tricritical point $K$, at which all
phases coalesce. As shown by Figs.~\ref{fig:bifurc} and \ref{fig:phase-B+}, three lines emanating from $K$} control the different phases: the
bifurcation line $\kappa_{b}(\theta)$, the coexistence line
$\kappa_{t}(\theta)$ and the first-order line $\kappa_{M}(\theta)$,
$\kappa_{b}(\theta)<\kappa_{t}(\theta)<\kappa_{M}(\theta)$. We
have obtained an exact expression for $\kappa_{b}(\theta)$ and
approximate analytical expressions for $\kappa_{t}(\theta)$ and
$\kappa_{M}(\theta)$ near the critical point. As shown in Fig.~\ref{fig:phase-B+}, their continuation far from $K$ describes
better the coexistence line $\kappa_{t}(\theta)$ than the first-order-line $\kappa_{M}(\theta)$. This is logical because they follow from a Landau-like expansion of the free energy around
the flat solution.

The above phase diagram is qualitatively similar to that found
numerically in a $2d$ version of the model, built on a hexagonal
lattice to model buckling and rippling in graphene \cite{RByP16}. It
is the qualitative shape of this phase diagram that explains the
emergence of the rippled to buckled transition when the system is
heated, recently observed in STM experiments \cite{schoelz15}. The key
point is the existence of values of the antiferromagnetic
parameter $\kappa$ (Region IIIa in Fig.~\ref{fig:bifurc}), for which the flat phase is locally stable at low temperature but becomes unstable and is \textit{replaced}
by a buckled phase when the temperature is increased. For sufficiently low initial temperature, we may prepare the string in a rippled flat profile that is a metastable
equilibrium state. As the temperature slowly increases
past the bifurcation line, the string suddenly jumps to and remains in a buckled state.

In light of the above discussion, it is tempting to conjecture that
the actual phase diagram of graphene is similar to the one found
here. The crux of the argument is the existence of some internal degrees of freedom analogous to pseudo-spins. For them: (i) their direct short-range interaction (of strength $\kappa$) favors rippling,
but (ii) their couplings to the elastic modes produce a
long-range interaction that favors buckling. It is this competition
that leads to a phase diagram like ours, in which there appear first order phase
transitions below some temperature. Then there appears a STM-like rippled-to-buckled transition as described in the previous paragraph.

In suspended graphene sheets, buckling instabilities may be due to residual stresses produced by the electron-phonon interaction \cite{ByR16}. This conclusion is based on a linear stability analysis of the flat configuration solution of saddle-point equations for phases in thermal equilibrium (first  deduced in \cite{gui14}). Whether buckling states bifurcate sub or supercritically from the flat membrane requires a study of not yet deduced small amplitude equations.

\acknowledgments This work has been supported by the Spanish
Mi\-nisterio de Econom\'\i a y Competitividad grants
MTM2014-56948-C2-2-P (MRG \& LLB) and FIS2014-53808-P (AP). MRG also
acknowledges support from MECD through the FPU program.

\appendix

\section{Stability of the low temperature
string  profiles}\label{sec:stability-low-temp-prof}

At low temperatures $\theta\to 0^{+}$, the Euler-Lagrange equation
becomes equivalent to
\begin{equation}\label{eq:EL-low-temp}
\frac{1}{\pi^{2}}u''+\sgn(u)\eta(|u|-2\kappa)=0, \, u(0)=u(1)=0.
\end{equation}
The solutions without internal nodes are \cite{Ruiz-Garcia}
\begin{equation}\label{eq:sol-low-temp}
  u^{(0)}(x)\!=\!\pm \!\left\{\begin{array}{ll}
                    \frac{\pi^{2}(1-2 x_{0}) x}{2}, & x<x_{0}, \\
                    2\kappa\!+\!\frac{\pi^{2}(x\!-\!x_{0})(1\!-\!x_{0}\!-\!x)}{2},
                                                    &
                                                      x_{0}\!<\!x\!<\!1\!-\!x_{0}, \\
                    \frac{\pi^{2}(1-2x_{0})(1-x)}{2}, & x>1-x_{0}.
\end{array}\right.
\end{equation}
The relation $u(x_0)=2\kappa$ produces the following equation for $x_{0}$:
\begin{equation}
\frac{\pi^{2}}{2}x_{0}(1-2x_{0})=2\kappa,
\end{equation}
Provided $\kappa<\kappa_{M}^{(0)}=\pi^{2}/32$, there are two solutions
$x_{0,j}$, $j=1,2$, given by Eq.~\eqref{eq:x0-values}, which are
symmetrical with respect to $1/4$, $x_{0,1}<1/4< x_{0,2}$. For
$\kappa>\kappa_{M}^{(0)}$, there are no buckled solutions and
$x_{0,1}=x_{0,2}=1/4$ if $\kappa=\kappa_{M}^{(0)}$.

Let $u^{(0)}(x)$ be one of these buckled stationary profiles
characterised by the sign in \eqref{eq:sol-low-temp} and the value of
$x_0$. To study its linear stability, we consider a small disturbance
from it, $u(x)=u^{(0)}(x)+\Delta u(x)$. According to the
  stability conditions described in Sec.~\ref{sec:Euler-Lagrange}, we
  have to solve the linear boundary value problem (BVP)
\begin{subequations}\label{eq:BVP}
\begin{equation}\label{eq:EL-low-temp-linear}
\frac{1}{\pi^{2}}\Delta u''+\delta(u^{(0)}(x)-2\kappa)\Delta u=0.
\end{equation}
\begin{equation}\label{eq:low-temp-stab-bc}
\qquad \Delta u(0)=\Delta u(a)=0, \quad a\leq 1.
\end{equation}
\end{subequations} Equation~\eqref{eq:EL-low-temp-linear}
  is the linearisation of Eq.~\eqref{eq:EL-low-temp} around
  \eqref{eq:sol-low-temp} (with positive sign). The profile
$u^{(0)}(x)$ is stable if, for any $a\leq 1$, $\Delta u(x)\equiv 0$ is
the unique solution of this BVP. On the contrary, if the
BVP has a non-trivial solution for some $a<1$, then $u^{(0)}(x)$ is
unstable.

Integrating \eqref{eq:EL-low-temp-linear} from $x_J-$ to $x_J+$ ($x_J$
is either $x_0$ or $1-x_0$), we find the jump conditions:
\begin{eqnarray}\label{eq:slope-jumps}
\Delta u'(x_{J}+)-\Delta u'(x_{J}-)=-\frac{2}{1-2x_{0}}\Delta
                   u(x_{J}).
\end{eqnarray}
As the solution of
\eqref{eq:EL-low-temp-linear}-\eqref{eq:low-temp-stab-bc} is unique up
to a multiplicative constant factor, we can fix the slope at $x=0$ to
be $\Delta u'(0)=1$ \cite{Gelfand-Fomin}. Then $\Delta u(0^{+})>0$. If
we find $\Delta u(1)<0$, then $\Delta u(a)=0$ at some intermediate
point $a\leq 1$ and the profile $u^{(0)}(x)$ is
unstable. 

 Equation~\eqref{eq:EL-low-temp-linear} tells us that
  $\Delta u(x)$ is composed of straight lines, with slope jumps at the
  points $x_{0}$ and $1-x_{0}$ determined by
  Eq.~\eqref{eq:slope-jumps}. Therefore,
\[
\Delta u=\left\{ \begin{array}{ll} x,& 0<x<x_0,\\
x_{0}+c_{1}(x-x_{0}),& x_{0}<x<1-x_{0},\\
x_{0}+c_{1}(1-2x_{0}) & \\
\quad\; +c_{2}(x-1+x_{0}),& 1-x_{0}<x<1.\\
\end{array}
\right.
\]
The jump conditions \eqref{eq:slope-jumps} readily yield
\begin{equation}\label{eq:c2-and-c3}
c_{1}=\frac{1-4x_{0}}{1-2x_{0}}, \qquad c_{2}=-1.
\end{equation}
Then,
\begin{equation}\label{eq:Delta-u-values}
\Delta u(1)=1-4x_{0}.
\end{equation}
Thus the stationary profile having $x_{0}>1/4$,
  corresponding to $x_{0,2}$ in Eq.~\eqref{eq:x0-values}, produces
$\Delta u(1)<0$ and it is unstable as explained above. For the other
stationary profile, corresponding to $x_{0,1}<1/4$, $\Delta u(x)$ is positive for
$0<x<1$ and the only solution of the BVP \eqref{eq:EL-low-temp-linear}
is $\Delta u=0$. Therefore this stationary profile is linearly stable.

\section{Effective Hamiltonian for the pseudo-spins}\label{sec:low-temp-spins}

We start by deriving the pseudo-spins' marginal probability
$\mathcal{P}(\bm{\sigma})$ by integrating the canonical distribution
$\mathcal{P}(\bm{u},\bm{p},\bm{\sigma})$ over the string degrees of
freedom. To do so, we rewrite Eq.~\eqref{H1} in matrix form,
\begin{equation}
 \mathcal{H}=\frac{1}{2m} \bm{p^T p}+ \frac{k}{2} \bm{u^T K
   u}-f\bm{u^T\sigma}+J \bm{ \sigma^T J \sigma},
\end{equation}
in which $(\bm{u},\bm{p},\bm{\sigma})$ are now column matrices
of dimension $N$, $(\bm{u^{T}},\bm{p^{T}},\bm{\sigma^{T}})$ are their
respective transpose matrices, and $\bm{J}$ and $\bm{K}$ are symmetric
matrices of dimension $N$, namely
\begin{equation}\label{K1}
 \bm{J} \!=\!\!  \begin{bmatrix}
                 0&\frac{1}{2}&&&&\\
		 \frac{1}{2}&0&\frac{1}{2}&&&\\
		 &\frac{1}{2}&0&\frac{1}{2}&&\\
		 &&& \ddots  \\
		 &&&\frac{1}{2}&0\!\!
                \end{bmatrix}\!,
\,
\bm{K} \!=\!\! \begin{bmatrix}
                 2&-1&&&&\\
		 -1&2&-1&&&\\
		 &-1&2&-1&&\\
		 &&& \ddots \\
		 &&&-1&2\!\!
                \end{bmatrix}\!.
\end{equation}
Also, we make the following change of variables,
$\bm{u}=\bm{v}+fk^{-1}\bm{\Lambda}\bm{\sigma}$, where $\bm{\Lambda}$ is the inverse of the matrix $\bm{K}$,
\begin{equation}\label{eq:Lambda-2}
\bm{\Lambda}_{ij}=\frac{1}{N+1} j (N-i+1) > 0, \, \forall i\ge j,
\, \bm{\Lambda}_{ij}=\bm{\Lambda}_{ji},
\end{equation}
see below for details on the derivation of the elements of
$\bm{\Lambda}$.

Interestingly, the variables $(\bm{v},\bm{p})$ and
$\bm{\sigma}$ become decoupled in the Hamiltonian, making it easy to
integrate the canonical distribution over $(\bm{v},\bm{p})$. The
result is
\begin{equation}
 \mathcal{P}(\bm\sigma)\! \propto\! e^{-\mathcal{H}_{\text{eff}}(\bm{\sigma})/\theta},\, \mathcal{H}_{\text{eff}}(\bm{\sigma})\!=\!\kappa \bm{\sigma^T\!
       J\sigma}\!-\!\frac{\pi^{2}}{2N^{2}}\bm{\sigma^T\! \bm{\Lambda}\sigma},
\label{eq1}
\end{equation}
which is Eq.~\eqref{eq:Prob-spins} of the main text.

Now, we derive the explicit expression of the elements of the matrix
$\bm{\Lambda}=\bm{K}^{-1}$. From equation Eq.~\eqref{K1}, we can directly
calculate the determinant of the matrix $\bm{K}_n$ ($\bm{K}$-matrix with
dimension $n$) as
\begin{subequations}
\begin{eqnarray}
 \det(\bm{K}_1)& = & 2, \quad
 \det(\bm{K}_2) = 3, \\
  \det(\bm{K}_n)& = &2 \det(\bm{K}_{n-1}) - \det(\bm{K}_{n-2}).
\end{eqnarray}
\end{subequations}
Hence,
\begin{equation}
\det(\bm{K}_n)=n+1.
\end{equation}
We take advantage of $\bm{K}$ being a symmetric matrix
$\bm{K}=\bm{K}^{\bm{T}}$, and impose  $i\ge j$ when calculating
$\bm{\Lambda}_{ij}$, which is also symmetric. Then, for dimension $N$
\begin{align}
\bm{\Lambda}_{ij} &=\frac{1}{N+1} (-1)^{i+j} \det(\bm{K}_{j-1})
                    (-1)^{i-j} \det(\bm{K}_{N-i}) \nonumber \\
   &=\frac{1}{N+1} j (N-i+1),
\end{align}
where we have made use of
\begin{equation}
\det \left( \begin{matrix}
                 \bm{A}& \bm{0}\\
		 \bm{B}&\bm{C}
\end{matrix}\right) =\det(\bm{A}) \det(\bm{C}),
\end{equation}
in which $\bm{A}$,  $\bm{B}$ and $\bm{C}$ are non-zero matrices and $\bm{0}$ the zero matrix.

\section{Effective Hamiltonian landscape}
\label{effective-H}

We want to characterise the $\mathcal{H}_{\text{eff}}$ landscape as
$\kappa$ is modified, where the phase space is formed by all possible
configurations of $\bm{\sigma}$.  For small enough $\kappa$, the
completely ferromagnetic configuration with all the pseudo-spins
pointing up (or down) minimises Eq.~\eqref{eq1}. On the other hand, as
$\kappa$ increases the configuration minimising Eq.~\eqref{eq1}
changes. Let us start from a completely ordered ferromagnetic
configuration $\bm{\sigma}_{\text{ferro}}$, in which $\sigma_i=+ 1$,
$\forall i$, and change the sign of $\sigma_l$, thereby obtaining the
configuration $R_{l}\bm{\sigma}_{\text{ferro}}$. The additional
contribution to the free energy is
 \begin{eqnarray}
\Delta\mathcal{H}_{\text{eff}}\!&\equiv&\! \mathcal{H}_{\text{eff}}(R_{l}\bm{\sigma}_{\text{ferro}})\!-\!\mathcal{H}_{\text{eff}}(\bm{\sigma}_{\text{ferro}})\nonumber\\
&=&\frac{ \pi^{2}}{2N^{2}}\sum\limits_{i\neq l}^{N}  \bm{\Lambda}_{l,i}-\kappa,
\label{16}
\end{eqnarray}
where
\begin{equation}\label{eq2}
 \sum\limits_{i\neq l}^{N}  \bm{\Lambda}_{l,i}=
 \frac{(N-1)(N+1-l)l}{2(N+1)}.
\end{equation}
This positive expression has a maximum at the centre, $l=(N+1)/2$, and
therefore $\Delta\mathcal{H}_{\text{eff}}$ is minimum when the
flipping pseudo-spins are those at the borders of the chain. This
suggests that, as $\kappa$ increases, the most probable (minimum free
energy) state will become antiferromagnetic at both boundaries while
remaining ferromagnetic in the bulk.

Now we can analyze the behavior of this global minimum with
increasing $\kappa$. In light of the discussion above, we restrict
ourselves to configurations in which $n_{a}$ consecutive
antiferromagnetic links have been created at each boundary, see
Fig.~\ref{fig:esq_na}. We denote by $\mathcal{H}_{\text{eff}}(n_a)$
the value of the effective Hamiltonian for such a configuration. Since
$n_{a}$ increases by two, we are interested in evaluating
$\mathcal{H}_{\text{eff}}(n_a)-\mathcal{H}_{\text{eff}}(n_a-2)$.
Using Eq.~\eqref{eq1}, taking into account the symmetries of $\bm K$
and $ \bm J$, and that we only have to take care of the terms that
change their sign from $\mathcal{H}_{\text{eff}}(n_a)$ to
$\mathcal{H}_{\text{eff}}(n_a-2)$, we get the expression
\begin{widetext}
\begin{eqnarray}
\mathcal{H}_{\text{eff}}(n_a)-\mathcal{H}_{\text{eff}}(n_a-2)&=&\frac{4 \pi^{2}}{N^{2}} \Bigg[ \sum\limits_{j=1}^{n_a-1} (-1)^{j} \frac{j(N-n_a+1)}{N+1}
+\sum\limits_{i=n_a+1}^{N-n_a}\frac{n_a(N-i+1)}{N+1}
                                                                 +\sum\limits_{i=N-n_a+2}^{N}(-1)^{i+1}\frac{(N-i+1)n_a}{N+1}\Bigg]\nonumber
  \\
&& -8\kappa.
\end{eqnarray}
After some simplifications,
\begin{equation}
\mathcal{H}_{\text{eff}}(n_a)-\mathcal{H}_{\text{eff}}(n_a-2)=\frac{2 \pi^2}{N^2}\left[-1+\left(1+N-2n_a\right)n_a\right]-8\kappa.
\label{Adiff_app}
\end{equation}
Iteration of this recurrence relation gives Eq.~\eqref{eq:Hna}.
\end{widetext}

\vspace{5ex}

\section{Stability of the phases}
\label{sec:stability}
Here we determine the stability of the different phases whose approximate profiles near bifurcation points, 
\begin{equation}\label{eq:eq-profile-lead}
u_{S}(x;C)=C \sin (\pi x),
\end{equation}
solve the Euler-Lagrange equations for the total free energy (see Sec.~\ref{sec:bifurcation}). Phase L (flat string profile) has $C=0$, whereas $C\neq 0$ for the buckled phases $B_\pm$. We shall calculate the total free energy for $u_S$ as a function of $C$ and determine whether it is a relative maximum or a minimum. The obtained stability results are consistent with the principle of exchange of stabilities in bifurcation theory \cite{Io90}.

The difference of free energies between the sinusoidal and the flat
profiles is given by
\begin{equation}\label{eq:free-energy-diff}
\Delta F(C;\kappa,\theta)\equiv \int_{0}^{1} dx \left[ f(u_{S},u'_{S};\kappa,\theta)-f_{\fla}(\kappa,\theta) \right].
\end{equation}
Note that $\Delta F$ is no longer a functional but a function of the
(unknown) amplitude $C$. To simplify our notation, we
  omit the dependence on $(\kappa,\theta)$ hereafter. Within the same
level of approximation as we have been working throughout, we have
\begin{eqnarray}\label{eq:deltaF-Bpm-expansion}
  \Delta F(C)&\sim&\int_{0}^{1}dx \left(\frac{1}{2}\dfb
    u_{S}^{2}+\frac{1}{4!}f_{4,b}u_{S}^{4}+\frac{1}{6!}f_{6,b}u_{S}^{6}\right)\!,
                                  \nonumber \\
&=&\frac{1}{4}\dfb C^{2}+\frac{1}{64}f_{4,b}C^{4}+\frac{1}{2304} f_{6,b}C^{6},
\end{eqnarray}
where $\dfb=f_{2}-f_{2,b}$, $f_{n,b}$ is the value of $f_{n}$ over the
bifurcation curve, as introduced in Sec.~\ref{sec:bifurcation}, and we
have neglected $O(C^{8})$ terms.  The equilibrium values of $C$, which
we denote by $C_{\eq}$, are found by seeking the extrema of
$\Delta F(C)$, see below.

Far from the critical point, consistently with the procedure for
solving perturbatively the Euler-Lagrange equation in
  Sec.~\ref{sec:bifurcation}, the term proportional to
  $C^{6}$ in Eq.~\eqref{eq:deltaF-Bpm-expansion} can be neglected.
Then, the non-vanishing values of $C_{\eq}$ obey
\begin{equation}\label{eq:C-sol}
C_{\eq}^{2}\sim -\frac{8\,\dfb}{f_{4b}}.
\end{equation}
 Note that $\dfb$ is of the order of $\epsilon^{2}$, cf Eqs.~\eqref{eq:dfb} and
  \eqref{eq:deltas}. Thus, $C_{\eq}$ is $O(\epsilon)$ and with
  the substitution $C_{\eq}=\epsilon A$, the above equation is
  completely equivalent to Eq.~\eqref{eq:amp1}. Insertion of $C_{\eq}$
  into Eq.~\eqref{eq:deltaF-Bpm-expansion} gives the free energy
  difference between the buckled and the flat phase,
\begin{equation}\label{eq:deltaF-Bpm-result-2}
\Delta F_{\eq}\sim -\frac{\dfb^{2}}{f_{4,b}}.
\end{equation}
which shows that the sign of $\Delta F_{\eq}$ is controlled by the
sign of $f_{4,b}$.

The stability of the phases can be further elucidated by looking at
the sign of the second derivative of $\Delta F$ with respect to
$C$, which is given by
\begin{equation}\label{eq:second-deriv}
\frac{\partial^{2}\Delta F}{\partial C^{2}}\sim\frac{\dfb}{2}+\frac{3}{16}f_{4b}C^{2}.
\end{equation}
Therefore,
\begin{equation}\label{eq:second-deriv-2}
\left.\frac{\partial^{2}\Delta F}{\partial C^{2}}\right|_{\eq}=-\dfb,
\end{equation}
and the stability is controlled by the sign of $\dfb$. We
  recall that $f_{4,b}$ vanishes at the critical point $\kappa$, and
  that $f_{4,b}>0$ ($f_{4,b}<0$) above (below)
  it. Then, above the critical point, the phase $B+$ bifurcates
inside the bifurcation line ($\dfb<0$) where the flat phase L becomes
unstable, and is thus stable: $\partial^{2}_{C}F(C)|_{B+}>0$ and,
consistently, $\left.\Delta F_{\eq}\right|_{B+}<0$.  To the right of
the critical point, the phase $B-$ emerges outside the bifurcation
line ($\dfb>0$), where the flat phase is stable, and is unstable:
$\partial^{2}_{C}F(C)|_{B-}<0$ and
$\left.\Delta F_{\eq}\right|_{B-}>0$. The phase $B-$ is indeed unstable
but it does not correspond to a (local) maximum of the free energy
functional, but to some kind of ``saddle point'' extremum that is
neither a minimum nor a maximum \cite{note3}.

In the vicinity of the tricritical point $K$, we have to keep the
$C^{6}$ terms, and substitute the coefficients of $\Delta F(C)$ with
their leading behaviors. With the same notation as before,
\begin{equation}\label{eq:deltaF-Bpm-vicinity}
\Delta F(C)\sim\frac{1}{4}\dfc C^{2}+\frac{1}{64}f_{4,c}^{(1)}C^{4}+\frac{1}{2304}f_{6,c}C^{6}.
\end{equation}
where
\begin{align}
\dfc=\epsilon^{4}\varphi_{4,c}, \quad
  f_{4,c}^{(1)}=\epsilon^{2}6\sqrt{3}\chi, \quad f_{6,c}=36,
\end{align}
and we have used Eqs.~\eqref{eq:deltas} (with
$\kappa_{2}=\theta_{2}=0$), \eqref{eq:tricrit_scales} and
\eqref{eq:phi4c}. Again, $C_{\eq}$ is found by looking for the extrema
of $\Delta F$, and $C_{\eq}=O(\epsilon)$. By introducing
$C_{\eq}=\epsilon A$, we have that $A$ is the solution of the
biquadratic equation~\eqref{eq:amp3}. Let us denote by $A_{\pm}^{2}$
the two solutions of Eq.~\eqref{eq:amp3}, with
$A_{+}^{2}>A_{-}^{2}$. As discussed in Sec.~\ref{sec:bifurcation}, (i)
above the critical point, $\chi>0$, it is only $A_{+}^{2}$ that makes
sense ($A_{-}^{2}<0$) and (ii) below the critical point, $\chi<0$,
both $A_{+}^{2}$ and $A_{-}^{2}$ are positive in a certain domain.

Again, the local stability of the phases is given by the second
derivative of $\Delta F$ at equilibrium. After a little algebra, one
gets the result
\begin{equation}\label{2-der-delta-F-Bpm}
\left.\frac{\partial^{2}\Delta F}{\partial C^{2}}\right|_{B\pm}\!\!\!\!=\pm\frac{\sqrt{3}}{4}\epsilon^{6}\tilde{A}_{\pm}^{2}\sqrt{9\chi^{2}-4\varphi_{4,c}}.
\end{equation}
Then the phase B+ is locally stable and the phase B- is unstable
within their respective domains of existence. Below the critical
point, we recall that the phase B+ exists for
$\kappa<\kappa_{M}(\theta)$, where $\kappa_{M}(\theta)$
  is the first-order line given by Eq.~\eqref{eq:kappaM}, whereas the
phase B- only exists between the bifurcation line and the first-order
line, $\kappa_{b}(\theta)<\kappa<\kappa_{M}(\theta)$. Over
$\kappa_{M}(\theta)$, both phases $B\pm$ merge, disappear and
$\partial^{2}\Delta F/\partial C^{2}|_{B\pm}=0$, because the argument
of the square root becomes equal to zero. Above the critical point,
only the plus sign is possible and Eq.~\eqref{2-der-delta-F-Bpm}
smoothly matches with Eq.~\eqref{eq:second-deriv-2}.

Let us focus on region III of the phase diagram in
  Fig.~\ref{fig:bifurc}, that is, between the bifurcation and the
  first-order line, $\kappa_{b}(\theta)<\kappa<\kappa_{M}(\theta)$.
Further analysis is necessary to find out which of the two locally
stable phases, the flat $L$ phase and the buckled $B+$ phase, gives
the absolute minimum of the free energy. The free energy
  difference $\Delta F$ is obtained by inserting $C_{\eq}=\epsilon A$
  in Eq.~\eqref{eq:deltaF-Bpm-vicinity}, which yields
\begin{equation}\label{eq:deltaF-sol-vicinity}
\Delta F_{B\pm}\!=\! \frac{\epsilon^{6}}{48}\tilde{A}_{\pm}^{2}\!
\left[8\varphi_{4,c}\!\pm 3\chi\sqrt{9\chi^{2}\!-\!4\varphi_{4,c}}\!-9\chi^{2}\right]\!.
\end{equation}
Recall that  $\chi<0$ below the critical point, and thus
$\sqrt{9\chi^2}=-3\chi$. Consistently with its unstable character, $\Delta F_{B-}\geq 0$, it
varies from $\Delta F_{B-}=0$ over the bifurcation line
$\kappa_{b}(\theta)$, at which $A_{-}$ vanishes, to the positive value
$\Delta F_{B}^{\max}=9\epsilon^{6}\tilde{A}^{2} \chi^{2}/48>0$ at the
first-order line $\kappa_{M}(\theta)$. On the other hand,
$\Delta F_{B+}<0$ at the bifurcation line, whereas
$\Delta F_{B+}=\Delta F_{B}^{\max}>0$ at the first order line because
the phases $B\pm$ merge. Thus, there must be a coexistence line at
which $\Delta F_{B+}$ vanishes and phases $B+$ and $L$ are equally
probable.  Equation~\eqref{eq:deltaF-sol-vicinity} determines the
condition $\varphi_{2,c}=27\chi^{2}/16$ or
\begin{equation}\label{eq:coex-line}
\kappa_{t}(\theta)\!=\!\kappa_{b}(\theta)+\frac{27\sqrt{3}}{96}(\theta\!-\!\theta_{c})^{2},
\, \theta<\theta_{c}, \ |\theta-\theta_{c}|\ll 1.
\end{equation}
For $\kappa_{b}(\theta)<\kappa<\kappa_{t}(\theta)$, the most stable
phase is B+, whereas the flat phase L is metastable; the situation is
just reversed in the region
$\kappa_{t}(\theta)<\kappa<\kappa_{M}(\theta)$.


\begin{thebibliography}{99}
\bibitem{note2}
We use L (from
level) instead of $F$, in order to avoid confusion in sentences such
as ``the free energy $F$ of the phase $F$''.
\bibitem{note1} In \cite{Ruiz-Garcia}, it
  was incorrectly stated that this configuration was metastable, but
  this does not affect reference \cite{Ruiz-Garcia} results and
  conclusions. The instability of the configuration for $x_{0,2}$ can
  also be shown from the pseudo-spins point of view, see discussion
  below.
\bibitem{note3} Note that the phase B+ cannot be produced by an
  expansion around the flat solution below the critical
  point, because its amplitude does not vanish at the bifurcation
  line. This is the reason why the solution obtained to the right of
  the critical point corresponds to the phase B-.
\end{thebibliography}
\end{document}